\documentclass[%
 reprint,
 amsmath,amssymb,
 pre,
nobalancelastpage,
]{revtex4-2}

\usepackage{graphicx}
\usepackage{dcolumn}
\usepackage{bm}
\usepackage{hyperref}
\hypersetup{
    colorlinks=true,
    linkcolor=blue,
    filecolor=magenta,
    urlcolor=cyan,
}

\usepackage[USenglish]{babel}
\usepackage[utf8]{inputenc}
\usepackage{mathtools}
\usepackage{amsfonts}
\usepackage{dcolumn}
\usepackage{booktabs}

\newcommand{\Appref}[1]{{Appx.~\ref{#1}}}
\newcommand{\Figref}[1]{{Fig.~\ref{#1}}}
\newcommand{\Eqnref}[1]{{Eq.~\ref{#1}}}
\newcommand{\Tblref}[1]{{Table~\ref{#1}}}

\newcommand{\leftnode}[0]{L}
\newcommand{\rightnode}[0]{R}

\usepackage[position=bottom]{subfig}
\begin{document}


\title{Mapping Flows on Bipartite Networks}

\author{Christopher Blöcker}
 \email{christopher.blocker@umu.se}
\author{Martin Rosvall}%
 \email{martin.rosvall@umu.se}
\affiliation{%
 Integrated Science Lab, Department of Physics \\
 Ume{\aa} University, SE-901 87 Ume{\aa}, Sweden
}%




\date{\today}

\begin{abstract}
Mapping network flows provides insight into the organization of networks, but even though many real-networks are bipartite, no method for mapping flows takes advantage of the bipartite structure.
What do we miss by discarding this information and how can we use it to understand the structure of bipartite networks better?
The map equation models network flows with a random walk and exploits the information-theoretic duality between compression and finding regularities to detect communities in networks.
However, it does not use the fact that random walks in bipartite networks alternate between node types, information worth $1$ bit.
To make some or all of this information available to the map equation, we developed a coding scheme that remembers node types at different rates.
We explored the community landscape of bipartite real-world networks from no node-type information to full node-type information and found that using node types at a higher rate generally leads to deeper community hierarchies and a higher resolution.
The corresponding compression of network flows exceeds the amount of extra information provided.
Consequently, taking advantage of the bipartite structure increases the resolution and reveals more network regularities.
\end{abstract}

\maketitle


\section{Introduction}
\label{sec:introduction}
Many networks are bipartite \cite{barber2007modularity, PhysRevE.76.036102, larremore2014efficiently}.
They model interactions between entities of different types: users watching movies, documents containing words, and animals eating plants.
Bipartite networks can also represent many-body interactions in hypergraphs: authors writing papers, proteins forming complexes, and people attending meetings.
Studying these networks with the naked eye is often infeasible because of their size and complexity.
Therefore, to carry out further analysis, we must simplify them.
We need to find coarse-grained descriptions that highlight their community structure \cite{fortunato2010community}.

Most community-detection methods are developed for unipartite networks, but can be used for bipartite networks as they are, either by running them on unipartite projections or by applying them directly to bipartite networks \cite{PhysRevE.93.032309, peixoto2013parsimonious}.
However, both these approaches have limitations.
First, unipartite projections of bipartite networks cannot preserve all the information that is encoded in the bipartite network such that significant structure is lost \cite{PhysRevE.76.036102}.
Second, applying unipartite methods directly to bipartite networks ignores the regularities of bipartite networks and does not take into account the fact that links only connect nodes of different types \cite{PhysRevE.102.032309}.
What do we miss by discarding this node-type information?
And how can we use it to understand the structure of bipartite networks better?

To explore the value of using bipartite information in community detection, we study the flow-based community-detection method Infomap \cite{mapequationAlgorithms}, which uses an information-theoretic objective function, known as the map equation \cite{Rosvall1118}, to exploit the duality between compression and finding regularities in data.
The map equation models network flows with random walks and relates the quality of a network partition to how well it compresses a modular description of the random walks.
Modules with long flow persistence, such as cliques or clique-like groups, achieve the best compression.
To derive a coding scheme, the map equation uses a hierarchical code that reflects the structure of the network partition.
However, this coding scheme is designed for unipartite networks and assumes that any pair of nodes can be connected and visited one after the other; it does not take advantage of the structural constraints in bipartite networks where links only connect nodes of different types and random walks must alternate between them.
Consequently, the map equation disregards bipartite information and provides suboptimal compression.

To address these issues, we developed a coding scheme that uses node-type information at different and adjustable rates.
For a node-type remembering rate of zero, we recover the standard map equation; a remembering rate of one leads to a fully bipartite map equation and higher compression.
Through intermediate rates, we can analyze how the community landscape changes with available node-type information.
We implemented the bipartite coding scheme in Infomap~\cite{bipartiteinfomap} and explored the community landscape of real-world networks from different domains.

In networks with community structure, we can compress flows beyond the extra information we make available through the coding scheme.
When we describe a network with all its nodes in one module, our coding scheme improves the compression by an amount equal to the entropy of the rate at which node types are used.
In hierarchical partitions, the compression improves proportionally to the available node-type information.
Generally, exploiting node types at higher rates increases the resolution and leads to deeper community structures with more and smaller modules, thus revealing more network regularities.

\section{The map equation framework}
To illustrate the duality between compression and finding regularities in network data, consider a communication game where the sender uses code words to update the receiver about the position of a random walker in a network.
We assume that the sender and receiver remember the current module but not the current node of the random walker.
The question is: how can we devise a modular coding scheme to minimize the average per-step description length, which we refer to as the \emph{code length}?

We start with all the nodes in one module and assign unique code words to the nodes based on their ergodic visit rates.
The sender needs to communicate exactly one code word per random-walker step to the receiver with this one-level approach.
According to Shannon's source-coding theorem \cite{Shannon1948}, the lower bound for the code length is the entropy of the node visit rates.

If the network has a community structure, we can achieve a lower code length with a two-level coding scheme: we partition the nodes into modules and define a separate codebook for each module.
This coding scheme uses unique code words within modules, allowing nodes in different modules to reuse short code words.
To describe transitions between modules for a uniquely decodable code, we introduce an index level codebook that assigns code words to modules, and add exit code words to each module codebook.
We can generalize this approach and reduce the code length further with a recursive code structure in multiple levels.

With a two-level approach, the sender communicates either one or three code words per random walker step.
For steps \emph{within} a module, the sender uses one code word from the current module codebook.
For transitions \emph{between} modules, the sender communicates three code words from three different codebooks:
\begin{enumerate}
  \item[(i)] the exit code word of the current module codebook,
  \item[(ii)] the entry code word of the new module from the index level codebook, and
  \item[(iii)] a node visit code word from the new module codebook.
\end{enumerate}
For a small example network (\Figref{fig:coding-a}), we illustrate the codebook structure for a two-level partition according to the map equation (\Figref{fig:coding-b}).

The map equation calculates the code length $L$ for a given partition $\mathsf{M}$ as the average of the module and index level code lengths, weighted by the fraction of time a random walker uses each of the corresponding codebooks in the limit,
\begin{equation}
  \label{eqn:standard}
  L \left(\mathsf{M}\right)
    = q H\left(\mathcal{Q}\right)
    + \sum_{\mathsf{m} \in \mathsf{M}} p_\mathsf{m} H\left(\mathcal{P}_\mathsf{m}\right).
\end{equation}
Here, $p_\mathsf{m} = q_\mathsf{m} + \sum_{n \in \mathsf{m}} p_n$ is the fraction of time the random walker uses the codebook for module $\mathsf{m}$, where $n \in \mathsf{m}$ are the nodes in $\mathsf{m}$, $p_n$ is the ergodic visit rate of node $n$, and $q_\mathsf{m}$ is the entry and exit rate of $\mathsf{m}$;
$q = \sum_{\mathsf{m} \in \mathsf{M}} q_\mathsf{m}$ is the rate at which the index level codebook is used; $\mathcal{Q} = \left\{ q_\mathsf{m} \,|\, \mathsf{m} \in \mathsf{M} \right\}$ is the set of module entry rates;
$\mathcal{P}_\mathsf{m} = \left\{ q_\mathsf{m} \right\} \cup \left\{ p_n \,|\, n \in \mathsf{m} \right\}$ is the set of node visit rates in module $\mathsf{m}$, including module exit;
and $H$ is the Shannon entropy.
We assume undirected networks and, therefore, entry and exit rates are the same.

To minimize the map equation, we need to make a tradeoff.
On the one hand, we want to keep modules small for short code words within modules.
On the other hand, we want to limit the number of modules for short code words at the index level.
Further, modules should have long flow persistence and cannot be too small; otherwise a random walker changes modules at a high rate and the sender is required to use the index level codebook frequently.
Under these restrictions, partitions with many links within modules and few links between modules give the best compression.

\section{The bipartite map equation}
Since the map equation was developed for unipartite networks, its coding scheme can describe transitions between any pair of nodes.
However, directly applying the map equation to bipartite networks leads to higher-than-necessary code lengths because transitions only happen between nodes of different types in bipartite networks.
For a more efficient coding scheme in bipartite networks, we consider the communication game again.
As before, the sender updates the receiver about the position of a random walker, but now both are aware of the bipartite network structure.

In a food web, for example, where herbivores are connected to plant species, random walks alternate between animal and plant nodes.
If the current node is an animal node, the random walker must step to a plant node next, and vice versa.
Therefore, we can use a bipartite coding scheme with two types of codebooks per module: one for animal-to-plant and one for plant-to-animal transitions.
Since both these codebooks only address half of the nodes on average, code words can be shorter.

To derive the code length of a bipartite coding scheme, we apply Bayes' rule to the standard map equation and obtain the bipartite map equation.
Let $\mathsf{M}_1$ be a partition with all nodes in one module and $\mathcal{P}_1$ be the set of ergodic node visit rates over two steps, that is, the visit rates we would obtain assuming a unipartite network.
The standard map equation calculates the entropy of the random process $X\colon$\emph{current node} from $\mathcal{P}_1$.
However, random walks on bipartite networks also provide information about a second process, namely, $Y\colon$\emph{current node type}.
In the bipartite map equation, we combine these two processes into one, $X|Y\colon$\emph{current node, given current node type}, and determine its entropy with Bayes' rule, $H\left(X|Y\right) = H\left(X\right) - H\left(Y\right) + H\left(Y|X\right)$.
We know that $H\left(Y\right) = 1\,$bit because the random walk alternates between nodes of different types and $H\left(Y|X\right) = 0\,$bits since the node fully determines the node type.
Let $\mathcal{P}^\leftnode$ and $\mathcal{P}^\rightnode$ be the sets of visit rates for left and right nodes, respectively, that is the two types of nodes in the bipartite network, given that the current node type is known.
Then, we can express $L\left(\mathsf{M}_1\right)$ in terms of $\mathcal{P}^\leftnode$ and $\mathcal{P}^\rightnode$,
\begin{equation}
  \label{eqn:one-level-entropy-rewrite}
  L\left(\mathsf{M}_1\right) = \underbracket{\,H\left(\mathcal{P}\right)\,}_{H\left(X\right)} = \underbracket{\ 1\ }_{H\left(Y\right)} + \underbracket{\,\frac{1}{2} H\left(\mathcal{P}^\leftnode\right) + \frac{1}{2} H\left(\mathcal{P}^\rightnode\right)\,}_{H\left(X|Y\right)},
\end{equation}
to show that providing the node type reduces the description of one-level partitions by $1\,$bit.

To generalize to two-level partitions, we plug this equation into \Eqnref{eqn:standard} and obtain the code length
\begin{equation}
  \label{eqn:hierarchical-rewrite}
  \begin{split}
  L\left(\mathsf{M}\right) = q \left( 1 + \frac{1}{2} H\left(Q^\leftnode\right) + \frac{1}{2} H\left(Q^\rightnode\right) \right) \hspace*{0.5em} \\
  + \sum_{\mathsf{m} \in \mathsf{M}} p_\mathsf{m} \left( 1 + \frac{1}{2} H\left(\mathcal{P}^\leftnode_\mathsf{m}\right) + \frac{1}{2} H\left(P^\rightnode_\mathsf{m} \right) \right),
  \end{split}
\end{equation}
where $\mathcal{Q}^\leftnode = \left\{q^\leftnode_\mathsf{m} \,|\, \mathsf{m} \in \mathsf{M} \right\}$ and $\mathcal{Q}^\rightnode = \left\{q^\rightnode_\mathsf{m} \,|\, \mathsf{m} \in \mathsf{M}\right\}$ are the sets of left and right module entry rates;
$\mathcal{P}^\leftnode_\mathsf{m} = \left\{q^\leftnode_\mathsf{m}\right\} \cup \left\{p_u \,|\, u \in \mathsf{m}^\leftnode\right\}$ and $\mathcal{P}^\rightnode_\mathsf{m} = \left\{q^\rightnode_\mathsf{m}\right\} \cup \left\{p_v \,|\, v \in \mathsf{m}^\rightnode\right\}$ are the sets of left and right node visit rates in module $\mathsf{m}$, including module exits;
$\mathsf{m}^\leftnode$ and $\mathsf{m}^\rightnode$ are the subsets of left and right nodes in $\mathsf{m}$;
and $p_u \in \mathcal{P}^\leftnode$ and $p_v \in \mathcal{P}^\rightnode$ are the visit rates for left nodes $u$ and right nodes $v$, respectively.

By separating the left and right visit rates in \Eqnref{eqn:hierarchical-rewrite}, we define the bipartite map equation:
\begin{equation}
  \label{eqn:bipartite}
  \begin{split}
  L_B\left(\mathsf{M}\right)
    = q^L H\left(\mathcal{Q}^L\right)
    + \sum_{\mathsf{m} \in \mathsf{M}} p^L_\mathsf{m} H\left(\mathcal{P}^L_\mathsf{m}\right) \hspace*{5pt} \\
    + \ q^R H\left(\mathcal{Q}^R\right)
    + \sum_{\mathsf{m} \in \mathsf{M}} p^R_\mathsf{m} H\left(\mathcal{P}^R_\mathsf{m}\right),
  \end{split}
\end{equation}
where $q^\leftnode = \sum_{m \in \mathsf{M}} q^\leftnode_m$ and $q^\rightnode = \sum_{m \in \mathsf{M}} q^\rightnode_m$ are the usage rates for left-to-right and right-to-left codebooks at index level;
$p^\leftnode_m = q^\leftnode_m + \sum_{u \in m^\leftnode} p_u$ and $p^\rightnode_m = q^\rightnode_m + \sum_{v \in m^\rightnode} p_v$ are the usage rates for left-to-right and right-to-left codebooks at the module level.
Thus, the bipartite map equation calculates the code length for a given partition that describes a joint clustering of left and right nodes in a bipartite network (detailed derivations in \Appref{appendix:derivation}).

The bipartite map equation changes the communication game.
As before, the sender uses one code word to encode transitions within modules and three code words for transitions between modules.
But now, both sender and receiver keep track of the current node type to choose the correct codebook -- left-to-right or right-to-left -- for their communication.

\pagebreak

\section{The bipartite map equation with varying node-type memory}
The map equation is about compression with constraints: compression is not the only goal.
The more we use the regularities in a network, the more we can compress its description.
But higher compression does not necessarily mean that we find network structures that allow us to understand the network better.

For example, consider a version of the coding game where sender and receiver remember the location of the random walker.
In this case, we would use a coding scheme with separate codebooks for each node with code words only for neighboring nodes.
This would allow us to encode the walker's path at the entropy rate of the corresponding Markov process \cite{Shannon1948} and provide a better compression than the map equation.
But then nodes would not have unique code words anymore and, even though the code is efficient, it would not capture the modular structure of the network.

The key is that the map equation forgets at which exact node a random walker is and only remembers the current module.
With the bipartite map equation, we relax this constraint by remembering node types.
However, in sparse bipartite networks, this comes close to remembering nodes and moves us towards encoding at the entropy rate of the Markov process without identifying modular structure.
Therefore, it is useful to look at using node-type information at intermediate rates.

\begin{figure*}
  \captionsetup[subfigure]{labelformat=empty}
  \centering
  \includegraphics[width=.8\textwidth]{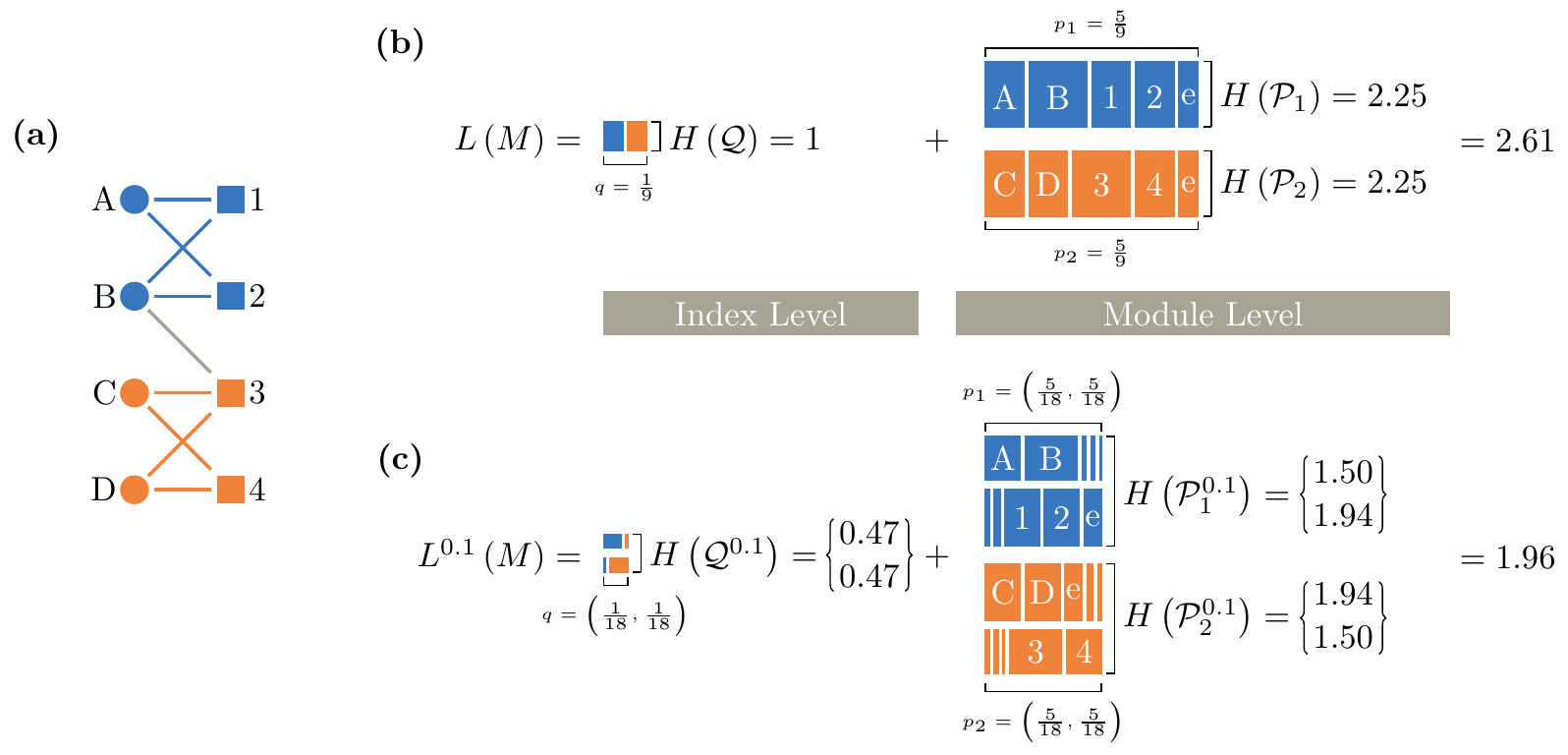}%
  \subfloat[\label{fig:coding-a}]{}%
  \subfloat[\label{fig:coding-b}]{}%
  \subfloat[\label{fig:coding-c}]{}%
  \caption{
    Graphical representation of the codebooks for the standard map equation and the bipartite map equation with $\alpha = 0.1$ in an unweighted example network where colors indicate modules.
    Block width corresponds to code word usage rate and block height to codebook entropy, a block's contribution to the map equation is its area.
    Letters in the blocks indicate which nodes they refer to, e stands for module exits.
    The horizontal gray bars show the contributions at index and module level.
    (a) The example network with color-coded modules.
    (b) The standard map equation calculates the code length as $2.61\,$bits.
    (c) Using node-type information worth $\mathcal{I} = 0.47\,$bits, the bipartite map equation with mixed node-type memory improves the compression by $0.65\,$bits to $1.96\,$bits.
  }
  \label{fig:coding}
\end{figure*}

In the bipartite map equation with varying node-type memory, node types are fuzzy.
While each node has a true type, either left or right, and the random walker alternates between types, we assume that we cannot determine types reliably.
We model this uncertainty by introducing a node-type flipping rate $\alpha$.
When we inspect a node, we observe its true type with probability $1-\alpha$, and the opposite type with probability $\alpha$.
Then, on average, nodes appear both left and right to a degree determined by $\alpha$.
Node-visit rates change accordingly and become mixed; we describe them as pairs of left and right flow: left nodes $u$ with visit rate $p_u$ have a mixed visit rate $p_u^\alpha = \left(\left(1-\alpha\right) p_u, \alpha p_u\right)$, and right nodes $v$ with visit rate $p_v$ have a mixed visit rate $p_v^\alpha = \left(\alpha p_v, \left(1-\alpha\right) p_v\right)$.

Using Bayes' rule again, we calculate the level of compression we can achieve when node types are fuzzy.
Let $\mathsf{M}_1$ be a partition with all the nodes in one module, $\mathcal{P}_1$ be the set of ergodic node visit rates, and $\mathcal{P}_1^\alpha = \left\{p_n^\alpha \,|\, n \in \mathsf{M}_1 \right\}$ be the set of mixed node visit rates.
The entropy of $Y\colon$\emph{current node type} is, as before, $1\,\text{bit}$ because we observe left and right nodes with probability $\frac{1}{2}$ each.
However, the entropy of $Y|X\colon$\emph{node type, given node} is now the entropy of the node-type flipping rate, $H\left(Y|X\right) = H_\alpha = H\left(1-\alpha,\alpha\right)$.
Overall, compared with the standard map equation, we can improve the compression by $1\,$bit, but node-type fuzziness increases the code length by $H_\alpha$, the entropy of the flipping rate,
\begin{equation}
  \label{eqn:one-level-entropy-rewrite-mixed}
  L\left(\mathsf{M}_1\right) = \underbracket{\,H\left(\mathcal{P}_1\right)\,}_{H\left(X\right)} = \underbracket{\ 1\ }_{H\left(Y\right)} - \underbracket{\,H_\alpha\,}_{H\left(Y|X\right)} + \underbracket{\,H\left(\mathcal{P}^\alpha_1\right)\,}_{H\left(X|Y\right)},
\end{equation}
where $H\left(\mathcal{P}_1^\alpha\right)$ is shorthand for the average component-wise entropies of the mixed node visit rates.

Plugging \Eqnref{eqn:one-level-entropy-rewrite-mixed} into the standard map equation gives us the generalisation to two-level partitions,
\begin{equation}
  \label{eqn:hierarchical-rewrite-mixed}
  \begin{split}
    L\left(\mathsf{M}\right) = q \left( 1 - H_\alpha + H \left( Q^\alpha \right) \right) \hspace*{5pt} \\
    + \sum_{\mathsf{m} \in \mathsf{M}} p_\mathsf{m} \left(1 - H_\alpha + H \left( P_\mathsf{m}^\alpha \right) \right).
  \end{split}
\end{equation}
We define the bipartite map equation with varying node-type memory:
\begin{equation}
  \label{eqn:mixed}
  L^\alpha\left(\mathsf{M}\right) = q^\alpha H\left(\mathcal{Q}^\alpha\right) + \sum_{\mathsf{m} \in \mathsf{M}} p^\alpha_\mathsf{m} H\left(\mathcal{P}_\mathsf{m}^\alpha\right),
\end{equation}
which measures the code length for a partition $\mathsf{M}$ and node-type flipping rate $\alpha$.
Figure \ref{fig:coding-c} illustrates how the codebook structure changes compared to the standard map equation (\Figref{fig:coding-b}) for a fixed value $\alpha$ in the same example network as before (\Figref{fig:coding-a}).
We can generalize the bipartite map equation with varying node type memory to more than two levels by recursively expanding the codebook structure within modules.
Then, each module within modules receives its own set of entry, node visit, and exit code words.

When node types are flipped at a rate of $\alpha = \frac{1}{2}$, nodes become left and right in equal parts.
With $H_\alpha = 1\,\text{bit}$, this means that there is maximum uncertainty about node types.
Ignoring node types in this way is equivalent to using the standard map equation.
The bipartite map equation is recovered for $\alpha = 0$ and $\alpha = 1$ because both values lead to $H_\alpha = 0$.
However, they have different interpretations.
For $\alpha = 0$, node types never flip and we can determine the true type of the nodes.
Under a flipping rate of $\alpha = 1$, node types always flip and we determine the opposite of the true node type.
This has no effect on the code length because it simply swaps the left and right entropy terms of the bipartite map equation.

Using the bipartite map equation with varying node-type memory, we are ready to answer the initial question: what more can we learn about a network by using node types in whole or in part?
Because it is more intuitive to think about how much we \emph{know} about node types than the probability of flipping them, we use entropy to connect these two quantities.
Flipping node types at rate $\alpha$ leads to an uncertainty of $H_\alpha$ about them.
Consequently, $\mathcal{I}\left(\alpha\right) = 1-H_\alpha$ is the available amount of information about node types, given that they are flipped at rate $\alpha$.
This formulation suggests an alternative interpretation of \Eqnref{eqn:one-level-entropy-rewrite-mixed}: we can reduce the code length of one-level partitions exactly by the amount of information that we have about node types.
To investigate by how much we can reduce the code length of two-level and hierarchical partitions, we have applied the bipartite map equation to real-world networks.

\section{Applying the bipartite map equation to real-world networks}
\label{sec:results}
We have implemented the bipartite map equation for two-level and hierarchical partitions in Infomap \cite{bipartiteinfomap}. The time complexity is the same as for standard Infomap, whose core algorithm is linear in the number of links.

We used the bipartite implementation to analyze the community landscape of 21 bipartite networks from different domains.
Our results show that the bipartite map equation uses node-type information effectively and improves the compression beyond the provided information.
The improved compression increases the resolution and lets us discover more regularities.

\begin{figure}
  \centering
  \subfloat[\label{fig:fonseca-ganade-0.30}]{%
    \includegraphics[width=.48\columnwidth]{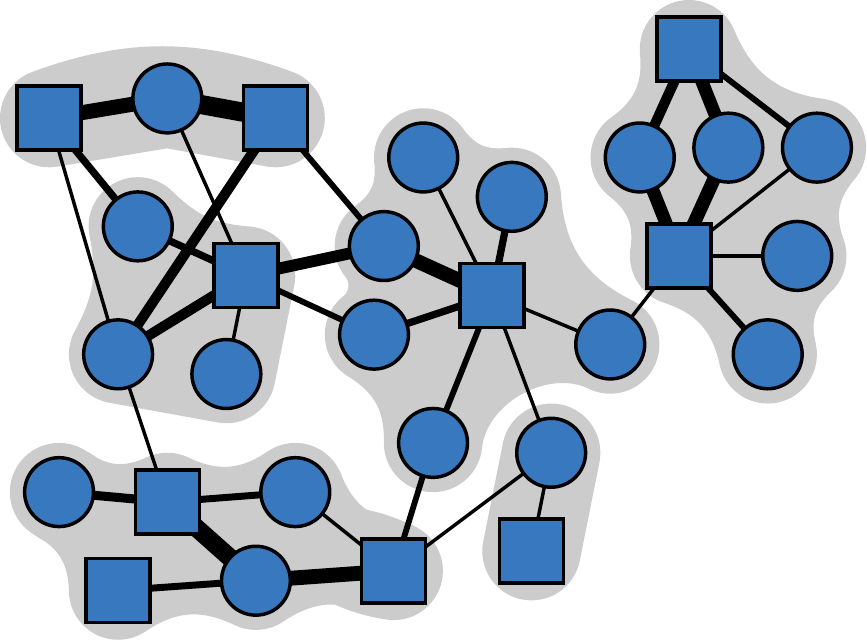}%
  }\hfill
  \subfloat[\label{fig:fonseca-ganade-0.84}]{%
    \includegraphics[width=.48\columnwidth]{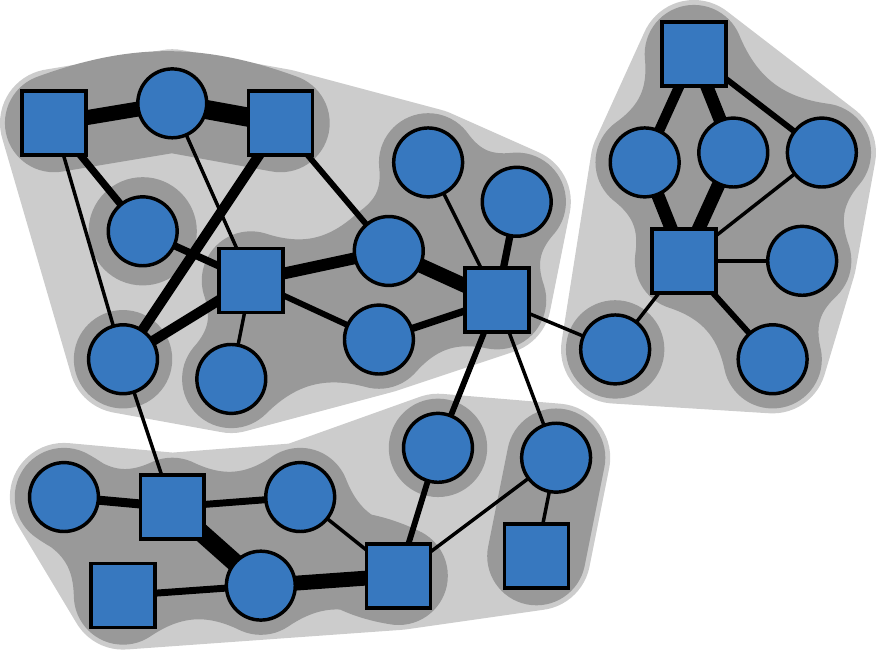}%
  }
  \caption{
    Community structure at different scales in the weighted Fonseca-Ganade plant-ant web \cite{10.2307/5880}. By providing more node-type information, we increase the resolution and detect finer modules on lower and coarser modules on higher levels in the community hierarchy.
    (a) Community structure for $\mathcal{I} = 0\,$bits ($\alpha = \frac{1}{2}$) with code length $2.24\,$bit and effective module size $5.27$.
    (b) Community structure for $\mathcal{I} = 0.65\,$bits ($\alpha = \frac{1}{6}$) with code length $1.5\,$bits and effective module size $4.35$.
  }
  \label{fig:new-structure}
\end{figure}

\subsection{Networks}
We selected 21 bipartite networks from different domains from the KONECT \cite{Kunegis:2013:KKN:2487788.2488173} and ICON \cite{ICON} databases and other sources \cite{10.1371/journal.pone.0171565, 10.1371/journal.pone.0148528}.
We preprocessed the networks with the python package NetworkX \cite{SciPyProceedings_11} and only kept their largest connected components.
The resulting networks ranged from a few dozen to millions of nodes and edges in size; their domain, number of left nodes $n_\leftnode$, number of right nodes $n_\rightnode$, and number of edges $m$ are listed in \Tblref{tbl:results}.
In weighted networks, marked with the superscript \textsuperscript{W}, the rate at which the random walker uses edges is proportional to their weight.
In all networks, left nodes represent subjects, such as users, documents, and animals, while right nodes represent objects that are acted upon, such as movies, words, and plants.

\subsection{Setup}
We explored the community landscape of our test networks from no information at $\mathcal{I} = 0\,$bits to full information at $\mathcal{I} = 1\,$bit with a step size of $0.05\,$bits.
For node-type information $\mathcal{I}$, we calculated the corresponding node-type flipping rate $\alpha$ numerically.
Because of its stochasticity, we ran Infomap 100 times for each network and value of $\alpha$, both with the flag \texttt{--two-level}, to search for two-level partitions, and without to search for hierarchical partitions.
Finally, for each $\alpha$, we selected the partitions with the best code length for further analysis.

\subsection{Structure and compression}
\begin{figure}
  \centering
  \subfloat[\label{fig:lvhk-sweep}]{%
    \includegraphics[width=.5\columnwidth]{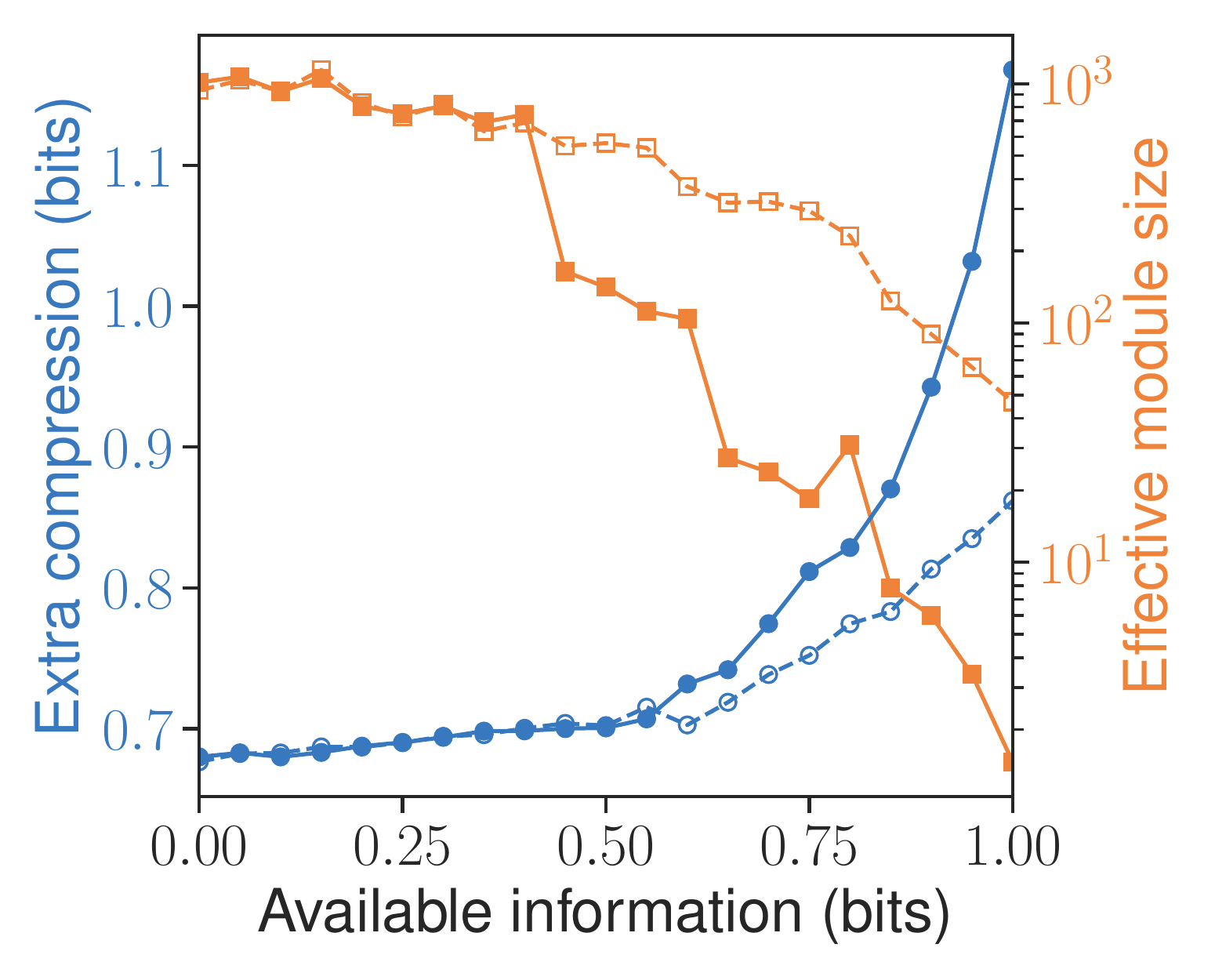}%
  }\hfill
  \subfloat[\label{fig:imdb-sweep}]{%
    \includegraphics[width=.5\columnwidth]{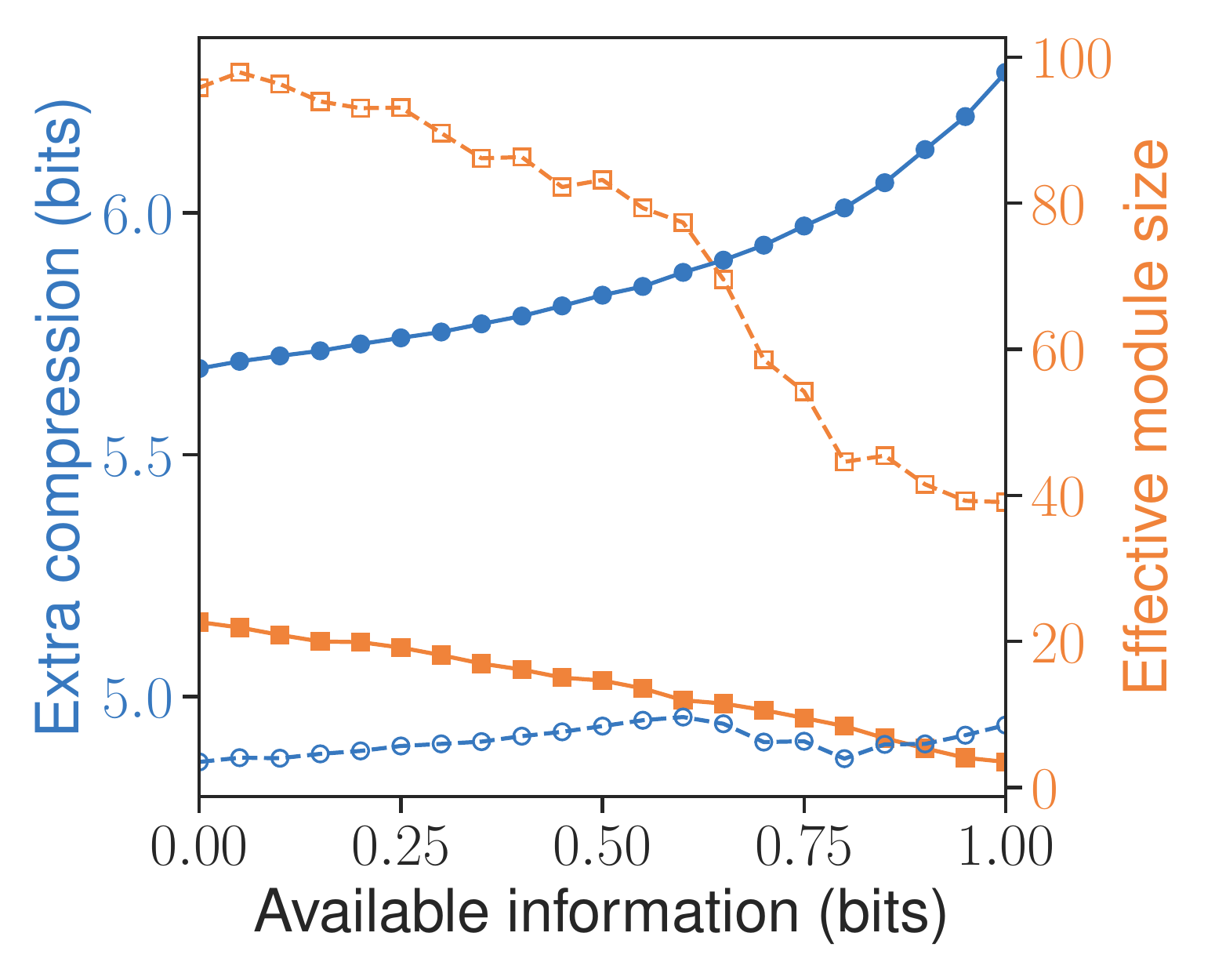}%
  }\hfill
  \subfloat[\label{fig:lastfm-song}]{%
    \includegraphics[width=.5\columnwidth]{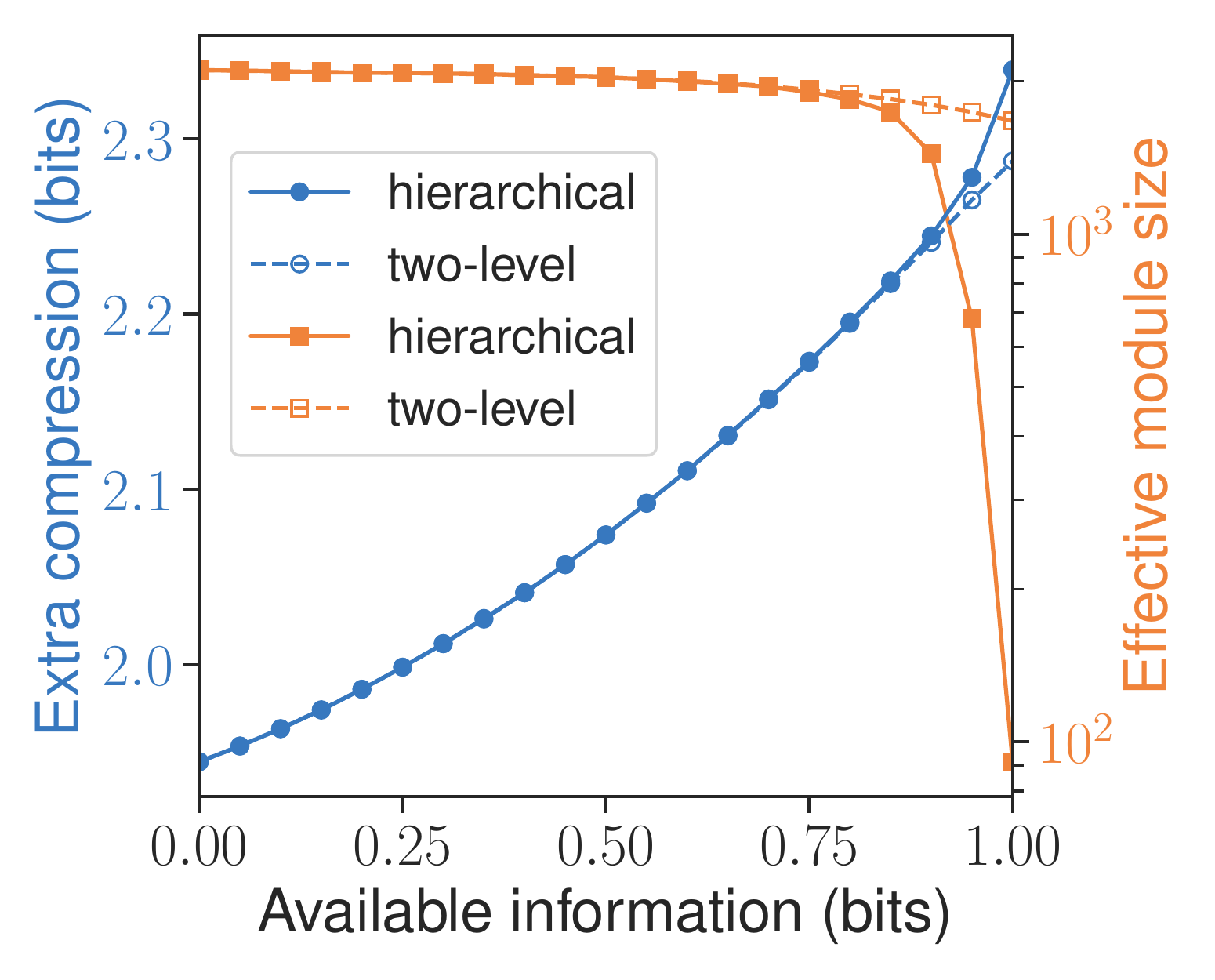}%
  }\hfill
  \subfloat[\label{fig:arroyo-goye-sweep}]{%
    \includegraphics[width=.5\columnwidth]{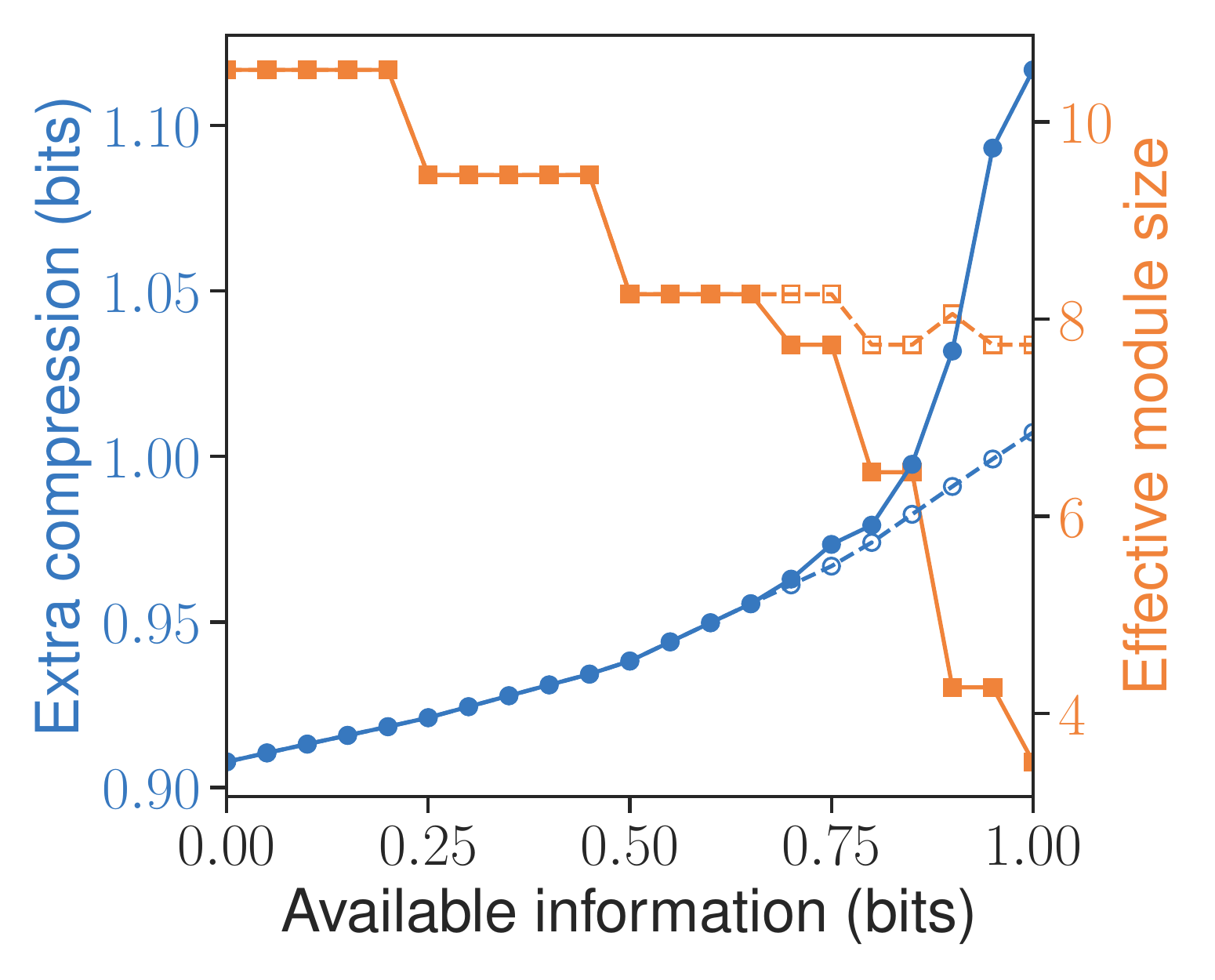}%
  }
  \caption{
  Compression and community resolution increase with available node-type information.
  The solid and dashed blue lines show the extra compression of the best hierarchical and two-level partitions, respectively.
  The solid and dashed orange lines show the effective module size of the best hierarchical and two-level partitions, respectively.
    (a) Las Vegas Hikers (LVHK) Meetup attendance.
    (b) IMDb actor-movie network.
    (c) Last.fm user-song network.
    (d) Arroyo Goye pollinator-plant web.
  }
  \label{fig:sweeps}
\end{figure}

\newcommand{\ww}[1]{#1\textsuperscript{W}}
\begin{table*}[ht]
  \caption{Properties of 21 bipartite test networks and their community landscape. The networks are sorted by number of edges, weighted networks are marked with the superscript \textsuperscript{W}. For each network and amount of node-type information, we ran Infomap 100 times and selected the hierarchical partitions with the best code length.}
  \label{tbl:results}
  \begin{tabular}{lcl c rrr c rrr c rrr}
    \toprule
    &&&&&&&& \multicolumn{3}{c}{Code length} && \multicolumn{3}{c}{Effective module size} \\
    \cmidrule{9-11}\cmidrule{13-15}
    Name & Ref & Domain & \hspace*{5pt} & $n_\leftnode$ & $n_\rightnode$ & m & \hspace*{5pt} & $\mathcal{I} = 0$ & $\mathcal{I} = 0.5$ & $\mathcal{I} = 1$ & \hspace*{5pt} & $\mathcal{I} = 0$ & $\mathcal{I} = 0.5$ & $\mathcal{I} = 1$ \\
    \midrule
    \ww{Wiktionary (en)}           & \cite{download.wikimedia.org}               & Authorship  &&    26,719 & 2,091,461 & 5,569,967 && 12.14 & 11.53 & 10.67 && 27{,}934   & 15{,}520   &  36   \\ 
    \ww{Last.fm user-song}         & \cite{Kunegis:2013:KKN:2487788.2488173}     & Interaction &&       992 & 1,084,620 & 4,413,834 && 12.33 & 11.70 & 10.94 &&  2{,}102   &  2{,}037   &  91   \\ 
    Wikipedia excellent            & \cite{download.wikimedia.org}               & Text        &&     2,780 &   273,959 & 2,941,902 && 13.64 & 13.15 & 11.80 && 68{,}944   &      145   &  24   \\ 
    \rule{0pt}{3ex}\noindent
    IMDb actor-movie               & \cite{IMDbActorMovie}                       & Affiliation &&   124,414 &   374,511 & 1,460,791 && 11.99 & 11.33 & 10.37 &&       23   &       15   &   3.5 \\ 
    Stack Overflow user-post       & \cite{konect:2017:stackoverflow}            & Rating      &&   524,670 &    80,492 & 1,280,982 && 11.83 & 11.04 &  9.99 &&       28   &       24   &   7.3 \\ 
    Reuters story-word             & \cite{Lewis:2004:RNB:1005332.1005345}       & Text        &&    19,757 &    38,677 &   978,446 && 13.44 & 12.89 & 11.94 &&  9{,}029   &  1{,}564   &   2.0 \\ 
    \rule{0pt}{3ex}\noindent
    \ww{Wiktionary (de)}           & \cite{download.wikimedia.org}               & Authorship  &&     5,354 &   144,710 &   686,661 && 11.23 & 10.65 &  9.82 &&  3{,}690   &  2{,}490   &   3.0 \\ 
    \ww{Linux kernel mailing list} & \cite{konect:2017:lkml_person-thread}       & Interaction &&    34,490 &   330,155 &   591,199 &&  9.61 &  9.02 &  8.22 &&      419   &      384   &  46   \\ 
    GitHub user-project            & \cite{GitHubContest}                        & Authorship  &&    39,845 &    99,907 &   417,361 && 11.24 & 10.52 &  9.21 &&       23   &       10   &   3.2 \\ 
    \rule{0pt}{3ex}\noindent
    YouTube user-group             & \cite{mislove-2007-socialnetworks}          & Affiliation &&    88,490 &    25,007 &   286,913 && 10.25 &  9.58 &  8.55 &&       48   &       29   &   5.4 \\ 
    APSMM conference               & \cite{10.1371/journal.pone.0148528}         & Social      &&    93,023 &        21 &   240,342 && 10.79 & 10.06 &  9.09 &&  6{,}742   &  3{,}663   & 108   \\ 
    LVHK Meetup                    & \cite{10.1371/journal.pone.0171565}         & Social      &&     6,061 &     5,096 &   127,033 && 11.58 & 11.06 & 10.09 &&  1{,}011   &      141   &   1.5 \\ 
    \rule{0pt}{3ex}\noindent
    PGHF Meetup                    & \cite{10.1371/journal.pone.0171565}         & Social      &&     4,989 &     4,611 &    39,501 && 10.90 & 10.26 &  9.26 &&       52   &       14   &   1.8 \\ 
    SIAM conference                & \cite{10.1371/journal.pone.0148528}         & Social      &&    10,018 &        19 &    15,533 &&  7.94 &  7.29 &  6.59 &&      525   &      427   &  89   \\ 
    NIPS conference                & \cite{10.1371/journal.pone.0148528}         & Social      &&     6,902 &        27 &    12,595 &&  8.14 &  7.38 &  6.74 &&      288   &      227   &  38   \\ 
    \rule{0pt}{3ex}\noindent
    \ww{UC Irvine forum}           & \cite{journals/socnet/Opsahl13}             & Social      &&       897 &       520 &     7,087 &&  8.16 &  7.61 &  6.93 &&       23   &       18   &   1.8 \\ 
    Norwegian directors            & \cite{SEIERSTAD201144}                      & Economic    &&       212 &       854 &     1,148 &&  3.83 &  3.07 &  2.03 &&        5.6 &        4.7 &   3.4 \\ 
    Virus-host interactome         & \cite{virus-host}                           & Biological  &&        41 &       288 &       433 &&  4.89 &  4.21 &  3.24 &&       17   &       13   &   5.7 \\ 
    \rule{0pt}{3ex}\noindent
    Scottish directors             & \cite{ScottHughes}                          & Economic    &&        86 &       131 &       348 &&  5.18 &  4.54 &  3.60 &&        6.8 &        5.1 &   2.1 \\ 
    \ww{Arroyo Goye}               & \cite{doi:10.1046/j.1461-0248.2003.00534.x} & Ecological  &&        27 &         8 &        41 &&  2.70 &  2.17 &  1.49 &&       11   &        8.2 &   3.5 \\ 
    \ww{Fonseca Ganade}            & \cite{10.2307/5880}                         & Ecological  &&        19 &        10 &        38 &&  2.24 &  1.68 &  1.06 &&        5.2 &        4.4 &   1.7 \\ 
    \bottomrule
  \end{tabular}
\end{table*}

We measured the extra compression provided by a partition $\mathsf{M}$ by using the corresponding one-level partition $\mathsf{M}_1$ as a baseline.
The one-level code length decreases by the amount of node-type information that is available (\Eqnref{eqn:one-level-entropy-rewrite-mixed}), specifically $L^\alpha\left(\mathsf{M}_1\right) = L^{0.5}\left(\mathsf{M}_1\right) - \mathcal{I}\left(\alpha\right)$, where $\mathcal{I}\left(\alpha\right) = 1 - H_\alpha$ is the node-type information when node-types are flipped at rate $\alpha$.
We define the extra compression of $\mathsf{M}$ as $L^\alpha\left(\mathsf{M}_1\right) - L^\alpha\left(\mathsf{M}\right) \geq 0$; it is always at least $0$ because Infomap returns the one-level partition when it does not find any partition with lower code length.
In partitions with more than one level, the extra compression depends on the codebook use rate, the total coding rate $q + \sum_{\mathsf{m} \in \mathsf{M}} p_\mathsf{m}$, and the amount of node-type information (\Eqnref{eqn:hierarchical-rewrite-mixed}).

To measure the resolution of the community detection, we use the effective module size as a proxy.
By only considering leaf modules -- those modules that contain nodes but have no sub-modules -- we can use the same measure for two-level and hierarchical solutions.
Let $S$ be the set of leaf module sizes in partition $\mathsf{M}$ where size refers to the number of nodes in a module.
Then the perplexity of the module sizes, $2^{H\left(S\right)}$ with $H\left(S\right) = - \sum_{s \in S} \frac{s}{\sum_{s' \in S} s'} \log \frac{s}{\sum_{s' \in S} s'}$, tells us the effective number of leaf modules, similar to how it can be used to calculate the effective number of sides of a (loaded) coin or die.
Combining the effective number of modules together with the number of nodes $N$ in the network, we calculate the effective module size as $\frac{N}{2^{H\left(S\right)}}$.

The effective community size and extra compression capture two significant patterns in the analyzed networks.

First, the resolution increases and we detect more communities on different scales when we use node types.
At lower levels in the community hierarchy, modules become more fine-grained, while on higher levels, they become more coarse.
For example, in the weighted Fonseca-Ganade plant-ant web, the bipartite map equation with $\mathcal{I} = 0.65\,$bits reveals hierarchically nested modules with smaller modules at the finest level (\Figref{fig:new-structure}).
With more node-type information, some nodes are assigned into singleton modules that form bridges between other modules.
The flow-persistence time is not long enough to include them in either of the other modules and, therefore, it is better to assign them to their own modules (\Figref{fig:new-structure}).
When we approach full node-type information at $\mathcal{I} = 1\,$bit, it can lead to so many small modules that no useful structure is detected anymore.
For example, leaf modules in the Las Vegas Hikers network (LVHK) contain only $1.5$ nodes on average (\Figref{fig:lvhk-sweep}).
In the IMDb actor-movie network, the effective module size decreases approximately linearly from $23$ at $\mathcal{I} = 0\,$bits to $3.5$ at $\mathcal{I} = 1\,$bit (\Figref{fig:imdb-sweep}).
In the Last.fm user-song network, the effective module size is around $2{,}000$ between $\mathcal{I} = 0\,$bits and $\mathcal{I} = 0.85\,$bits but then drops sharply and is $91$ for $\mathcal{I} = 1\,$bit (\Figref{fig:lastfm-song}).
We see a similar behavior in all the networks we analyzed (\Tblref{tbl:results}), both for hierarchical and two-level partitions, with the difference being that sharp drops in module size are less common in two-level partitions (\Figref{fig:sweeps}, \Appref{appendix:plots}).
However, as leaf modules become smaller, the community hierarchy becomes deeper such that higher levels still contain significant structures.

Second, the compression improves by more than the amount of node-type information we provide.
With the duality between compression and finding regularities in data, the bipartite map equation detects more structure in the bipartite networks.
Because the entropy function is non-linear, the extra compression generally increases faster with more available information.
For example, in the IMDb actor-movie network, when the code length decreases from $11.99$\,bits at $\mathcal{I} = 0\,$bits to $10.37$\,bits at $\mathcal{I} = 1\,$bit, the extra compression improves from $5.7\,$bits to $6.3\,$bits, and the rate of improvement increases closer to full node-type information (\Tblref{tbl:results}, \Figref{fig:imdb-sweep}).
In the Arroyo Goye pollinator-plant web and the LVHK network, the compression improves slowly at first, but faster once more regularities can be detected above $\mathcal{I} = 0.5\,$bits (\Figref{fig:lvhk-sweep}, \Figref{fig:arroyo-goye-sweep}).
However, since we ran Infomap independently for each $\alpha$, the extra compression sometimes decreased with more information (\Figref{fig:lvhk-sweep}, \Figref{fig:imdb-sweep}, \Appref{appendix:plots}).
In these cases, Infomap's stochastic search algorithm did not find partitions that would have led to an increase in extra compression.
For example, using the same partition over the whole range of $\alpha$ guarantees a monotonic increase in all networks.
Nevertheless, by providing more node-type information, the regularizing effect of the standard map equation decreases and the compression generally increases.

Higher resolution and further compression also result from using shorter Markov times~\cite{delvenne2010stability,schaub2012encoding}, but the map equation for varying Markov times~\cite{PhysRevE.93.032309} and the bipartite map equation work in different ways. Short Markov times correspond to a lazy random walker on a modified network with strong self-links. With fewer steps between nodes, cheaper transitions between communities shift the optimal solution to smaller communities with shorter average code lengths. Instead, the bipartite map equation transforms node type information into compression with smaller node-type specific codebooks. Cheaper transitions between communities then shift the optimal solution to smaller communities and result in extra compression.

\section{Conclusion}
\label{sec:conclusion}
We have extended the map equation framework for finding modules in network flows to use node-type information encoded in bipartite networks.
Applied to 21 real-world networks, the bipartite map equation implemented in the search algorithm Infomap detects more, smaller communities at lower levels of the community hierarchy and fewer, larger modules at higher levels.
The community-detection resolution increases because the bipartite map equation's coding scheme exploits the alternating trajectories of random walks and compresses the description of network flows beyond the provided node-type information.
In between ignoring and making full use of the node-type information, the bipartite map equation can use the node-type information at intermediate rates, offering a principled way to explore communities at higher resolution in bipartite networks.

\begin{acknowledgements}
This work was partially supported by the Wallenberg AI, Autonomous Systems and Software Program (\href{http://wasp-sweden.org}{WASP}) funded by the Knut and Alice Wallenberg Foundation.
We would like to thank Leto Peel, Vincenzo Nicosia, and Jelena Smiljani\'c for discussions that helped to improve this paper.
Martin Rosvall was supported by the Swedish Research Council, Grant No.\ 2016-00796.
\end{acknowledgements}

\providecommand{\noopsort}[1]{}\providecommand{\singleletter}[1]{#1}%

\clearpage
\onecolumngrid
\appendix

\clearpage
\section{Derivation of the bipartite map equation}
\label{appendix:derivation}
\noindent
Consider an undirected, weighted bipartite graph $G = \left( N_\leftnode, N_\rightnode, E, \delta \right)$ with left nodes $N_\leftnode$, right nodes $N_\rightnode$, edges $E \subseteq N_\leftnode \times N_\rightnode$, and edge weights $\delta \colon E \to \mathbb{R}$.
Let $\mathcal{P}^\leftnode = \left\{ p_u \,|\, u \in N_\leftnode \right\}$ and $\mathcal{P}^\rightnode = \left\{ p_v \,|\, v \in N_\rightnode \right\}$ be the left and right node visit rates.
Since the graph is undirected, we can calculate the visit rates directly by $p_u = \frac{\sum_{v \in N_\rightnode} \delta\left(\left(u,v\right)\right)}{\Delta\left(G\right)}$ for left nodes $u$ and $p_v = \frac{\sum_{u \in N_\leftnode} \delta\left(\left(u,v\right)\right)}{\Delta\left(G\right)}$ for right nodes $v$, where $\Delta\left(G\right) = \sum_{e \in E} \delta\left(e\right)$ is the total edge weight in $G$.
The visit rate of a disconnected node is $0$, but we exclude such nodes from our considerations because they could be assigned to any module without affecting the code length.
Since the graph is bipartite, both $\mathcal{P}^\leftnode$ and $\mathcal{P}^\rightnode$ sum to $1$, that is, $\sum_{p_u \in \mathcal{P}^\leftnode} p_u = 1$ and $\sum_{p_v \in \mathcal{P}^\rightnode} p_v = 1$.

\noindent
Let $N = N_L \cup N_R$ be the set of all nodes and $\mathcal{P}$ be the set of ergodic visit rates over two steps, that is, the visit rates we would obtain when we assume a unipartite network.
For distinction between node types, we use $u$ to refer to left nodes, $v$ to refer to right nodes, and $n$ when we talk about both types in combination.
We denote left and right visit rates by $p_u$ and $p_v$, respectively, and ergodic visit rates over two steps by $\overline{p}_n$.
Since the graph is bipartite, the total weight of edges incident to left nodes is equal to the total weight of edges incident to right nodes and, therefore, the ergodic visit rate over two steps for a node $n$ is $\overline{p}_n = \frac{p_n}{2}$.
Then the set of ergodic visit rates over two steps is connected to the left and right visit rates by
\begin{equation}
  \label{eqn:derivation-visit-rates}
  \mathcal{P} = \left\{ \frac{p_u}{2} \,|\, p_u \in \mathcal{P}^\leftnode \right\} \cup \left\{ \frac{p_v}{2} \,|\, p_v \in \mathcal{P}^\rightnode \right\}.
\end{equation}

\noindent
Let $\mathsf{M}$ be a partition of the nodes into modules.
The standard map equation calculates the code length of $\mathsf{M}$ as the average of the module and index level code lengths, weighted by the fraction of time a random walker uses each of the codebooks,
\begin{equation}
  \label{eqn:derivation-standard}
  L\left(\mathsf{M}\right) = q H\left(\mathcal{Q}\right) + \sum_{\mathsf{m} \in \mathsf{M}} p_\mathsf{m} H\left(\mathcal{P}_\mathsf{m}\right).
\end{equation}
Here, $p_\mathsf{m} = q_\mathsf{m} + \sum_{n \in \mathsf{m}} \overline{p}_n$ is the fraction of time the random walker uses the codebook for module $\mathsf{m}$ and $n \in \mathsf{m}$ are the nodes in $\mathsf{m}$, and $q_\mathsf{m}$ is the entry and exit rate of $\mathsf{m}$; $q = \sum_{\mathsf{m} \in \mathsf{M}} q_\mathsf{m}$ is the rate at which the index level codebook is used.
$\mathcal{Q} = \left\{ q_\mathsf{m} \,|\, \mathsf{m} \in \mathsf{M} \right\}$ is the set of module entry rates, $\mathcal{P}_\mathsf{m} = \left\{ q_\mathsf{m} \right\} \cup \left\{ \overline{p}_n \,|\, n \in \mathsf{m} \right\}$ is the set of node visit rates in module $m$, including module exit, and $H$ is the Shannon entropy.
Note than entry and exit rates are identical since the network is undirected.

\noindent
Let $\mathsf{M}_1$ be a \emph{one-level} partition with all nodes in the same module.
Then the code length according to the standard map equation is
\begin{align}
L \left( \mathsf{M}_1 \right) = H \left( \mathcal{P} \right)
  &=\, - \sum_{\overline{p}_n \in \mathcal{P}} \overline{p}_n \log_2 \overline{p}_n \nonumber \\
  &=\, - \sum_{\overline{p}_n \in \mathcal{P}} \overline{p}_n \left( \log_2 2\overline{p}_n - 1 \right) \nonumber \\
  &=\, - \sum_{\overline{p}_n \in \mathcal{P}} \overline{p}_n \log_2 2\overline{p}_n + \sum_{\overline{p}_n \in P} \overline{p}_n \nonumber \\
  &=\, 1 - \frac{1}{2} \sum_{\overline{p}_n \in \mathcal{P}} 2\overline{p}_n \log_2 2\overline{p}_n \nonumber \\
  &\overset{\ref{eqn:derivation-visit-rates}}{=} 1 - \frac{1}{2} \sum_{p_u \in \mathcal{P}^\leftnode} p_u \log_2 p_u - \frac{1}{2} \sum_{p_v \in \mathcal{P}^\rightnode} p_v \log_2 p_v \nonumber \\
  &=\, 1 + \frac{1}{2} H \left( \mathcal{P}^\leftnode \right) + \frac{1}{2} H \left( \mathcal{P}^\rightnode \right). \label{eqn:derivation-one-level-rewrite}
\end{align}
To generalize, we plug \Eqnref{eqn:derivation-one-level-rewrite} into \Eqnref{eqn:derivation-standard},
\begin{equation}
  \label{eqn:derivation-two-level-rewrite}
  L\left(\mathsf{M}\right)
    = q \left( 1 + \frac{1}{2} H\left(Q^\leftnode\right) + \frac{1}{2} H\left(Q^\rightnode\right) \right)
    + \sum_{\mathsf{m} \in \mathsf{M}} p_\mathsf{m} \left( 1 + \frac{1}{2} H\left(\mathcal{P}^\leftnode_\mathsf{m}\right) + \frac{1}{2} H\left(P^\rightnode_\mathsf{m} \right) \right).
\end{equation}
Here, $\mathcal{Q}^\leftnode = \left\{q^\leftnode_\mathsf{m} \,|\, \mathsf{m} \in \mathsf{M} \right\}$ and $\mathcal{Q}^\rightnode = \left\{q^\rightnode_\mathsf{m} \,|\, \mathsf{m} \in \mathsf{M}\right\}$ are the sets of left and right module entry rates; $\mathcal{P}^\leftnode_\mathsf{m} = \left\{q^\leftnode_\mathsf{m}\right\} \cup \left\{p_u \,|\, u \in \mathsf{m}^\leftnode\right\}$ and $\mathcal{P}^\rightnode_\mathsf{m} = \left\{q^\rightnode_\mathsf{m}\right\} \cup \left\{p_v \,|\, v \in \mathsf{m}^\rightnode\right\}$ are the sets of left and right node visit rates in module $\mathsf{m}$, including module exits.
Further, $\mathsf{m}^\leftnode$ and $\mathsf{m}^\rightnode$ are the subsets of left and right nodes in $\mathsf{m}$.

\noindent
Based on \Eqnref{eqn:derivation-two-level-rewrite}, we define the bipartite map equation,
\begin{equation}
  \label{eqn:derivation-bipartite}
  L_B\left(\mathsf{M}\right)
    = q^L H\left(\mathcal{Q}^L\right)
    + \sum_{\mathsf{m} \in \mathsf{M}} p^L_\mathsf{m} H\left(\mathcal{P}^L_\mathsf{m}\right)
    + \ q^R H\left(\mathcal{Q}^R\right)
    + \sum_{\mathsf{m} \in \mathsf{M}} p^R_\mathsf{m} H\left(\mathcal{P}^R_\mathsf{m}\right).
\end{equation}
Here, $q^\leftnode = \sum_{\mathsf{m} \in \mathsf{M}} q^\leftnode_\mathsf{m}$ and $q^\rightnode = \sum_{\mathsf{m} \in \mathsf{M}} q^\rightnode_\mathsf{m}$ are the usage rates for left-to-right and right-to-left codebooks at index level; $p^\leftnode_\mathsf{m} = q^\leftnode_\mathsf{m} + \sum_{u \in \mathsf{m}^\leftnode} p_u$ and $p^\rightnode_\mathsf{m} = q^\rightnode_\mathsf{m} + \sum_{v \in \mathsf{m}^\rightnode} p_v$ are the usage rates for left-to-right and right-to-left codebooks at module level, respectively.
As the total weight of edges incident to left nodes is equal to the total weight of edges incident to right nodes, we have $q^\leftnode = q^\rightnode = \frac{q}{2}$ and $p^\leftnode_\mathsf{m} = p^\rightnode_\mathsf{m} = \frac{p_\mathsf{m}}{2}$ for all $\mathsf{m}$.

\noindent
Consider again $\mathcal{P}$, the set of ergodic node visit rates over two steps and let $\alpha \in \left[0,1\right] \subset \mathbb{R}$.
For better readability and because specific nodes are not important, we refer to the visit rates over two steps simply as $p$ in the following.
Further, we use $H_\alpha = H\left(1-\alpha, \alpha\right)$ as shorthand for the entropy of $\alpha$.
We can then rewrite $H\left(\mathcal{P}\right)$,
\begin{align}
  H \left(\mathcal{P}\right) \nonumber
    &= -\sum_{p \in \mathcal{P}} p \log_2 p \nonumber \\
    &= \left(\left(1-\alpha\right) + \alpha \right) \left(-\sum_{p \in \mathcal{P}} p \log_2 p \right) \nonumber \\
    &= \left(1-\alpha\right) \left(-\sum_{p \in \mathcal{P}} p \log_2 p \right) + \alpha \left(-\sum_{p \in \mathcal{P}} p \log_2 p \right) \nonumber \\
    &= -\sum_{p \in \mathcal{P}} \left(\left(1-\alpha\right) p \right) \log_2 p - \sum_{p \in \mathcal{P}} \alpha p \log_2 p \nonumber \\
    &= -\sum_{p \in \mathcal{P}} \left(\left(1-\alpha\right) p \right) \log_2 \frac{\left(1-\alpha\right) p}{1-\alpha} - \sum_{p \in \mathcal{P}} \alpha p \log_2 \frac{\alpha p}{\alpha} \nonumber \\
    &= -\left(\sum_{p \in \mathcal{P}} \left(\left(1-\alpha\right) p \right) \log_2 \left(\left(1-\alpha\right) p \right) - \left(\left(1-\alpha\right) p \right) \log_2 \left(1-\alpha\right) \right) - \left(\sum_{p \in \mathcal{P}} \alpha p \log_2 \alpha p - \alpha p \log_2 \alpha \right) \nonumber \\
    &= -\sum_{p \in \mathcal{P}} \left(\left(1-\alpha\right) p \right) \log_2 \left(\left(1-\alpha\right) p \right) + \sum_{p \in \mathcal{P}} \left(\left(1-\alpha\right) p \right) \log_2 \left(1-\alpha\right) - \sum_{p \in \mathcal{P}} \alpha p \log_2 \alpha p + \sum_{p \in \mathcal{P}} \alpha p \log_2 \alpha \nonumber \\
    &= \left(1-\alpha\right) \log_2 \left(1-\alpha\right) + \alpha \log_2 \alpha - \sum_{p \in \mathcal{P}} \left(\left(1-\alpha\right) p \right) \log_2 \left(\left(1-\alpha\right) p \right) - \sum_{p \in \mathcal{P}} \alpha p \log_2 \alpha p \nonumber \\
    &= -H_\alpha - \sum_{p \in \mathcal{P}} \left(\left(1-\alpha\right) p \right) \log_2 \left(\left(1-\alpha\right) p \right) - \sum_{p \in \mathcal{P}} \alpha p \log_2 \alpha p \label{eqn:derivation-alpha-fanout}
\end{align}

\noindent
With \Eqnref{eqn:derivation-one-level-rewrite} and \Eqnref{eqn:derivation-alpha-fanout}, we rewrite the code length of the one-level partition,
\begin{align}
  L\left(\mathsf{M}_1\right) = H\left(\mathcal{P}\right)
    & \overset{\ref{eqn:derivation-one-level-rewrite}}{=} 1 + \frac{1}{2} H\left(\mathcal{P}^\leftnode\right) + \frac{1}{2} H\left(\mathcal{P}^\rightnode\right) \nonumber \\
    & \overset{\ref{eqn:derivation-alpha-fanout}}{=} 1 + \frac{1}{2} \left(-H_\alpha - \sum_{p_u \in \mathcal{P}^\leftnode} \left(\left(1-\alpha\right) p_u \right) \log_2 \left(\left(1-\alpha\right) p_u \right) - \sum_{p_u \in \mathcal{P}^\leftnode} \alpha p_u \log_2 \alpha p_u \right) \nonumber \\
    & \hspace*{20.5pt} + \frac{1}{2} \left(-H_\alpha - \sum_{p_v \in \mathcal{P}^\rightnode} \left(\left(1-\alpha\right) p_v \right) \log_2 \left(\left(1-\alpha\right) p_v \right) - \sum_{p_v \in \mathcal{P}^\rightnode} \alpha p_v \log_2 \alpha p_v \right) \nonumber \\
    &= 1 - H_\alpha + \frac{1}{2} \left(-\sum_{p_u \in \mathcal{P}^\leftnode} \left(\left(1-\alpha\right) p_u \right) \log_2 \left(\left(1-\alpha\right) p_u \right) - \sum_{p_v \in \mathcal{P}^\rightnode} \alpha p_v \log_2 \alpha p_v \right) \nonumber \\
    & \hspace*{44.5pt} + \frac{1}{2} \left(-\sum_{p_v \in \mathcal{P}^\rightnode} \left(\left(1-\alpha\right) p_v \right) \log_2 \left(\left(1-\alpha\right) p_v \right) - \sum_{p_u \in \mathcal{P}^\leftnode} \alpha p_u \log_2 \alpha p_u \right) \nonumber \\
    &= 1-H_\alpha + \frac{1}{2} H\left(\mathcal{R}^\alpha\right) + \frac{1}{2} H\left(\mathcal{R}^{1-\alpha}\right) \label{eqn:derivation-one-level-rewrite-alpha},
\end{align}
where we define mixed node visit rates,
\begin{equation}
  \label{eqn:derivation-r-alpha}
  \mathcal{R}^\alpha = \left\{ \left(1-\alpha\right) p_u \,|\, p_u \in \mathcal{P}^\leftnode \right\} \cup \left\{ \alpha p_v \,|\, p_v \in \mathcal{P}^\rightnode \right\}.
\end{equation}
For values of $\alpha = 0$, $\alpha = 1$, and $\alpha = \frac{1}{2}$, we retrieve the original definitions of $\mathcal{P}^\leftnode$, $\mathcal{P}^\rightnode$, and $\mathcal{P}$, respectively, from \Eqnref{eqn:derivation-r-alpha}.
Again, to generalize, we plug \Eqnref{eqn:derivation-one-level-rewrite-alpha} into \Eqnref{eqn:derivation-standard},
\begin{equation}
  \label{eqn:derivation-two-level-rewrite-alpha}
  L\left(\mathsf{M}\right)
    = q \left(1 - H_\alpha + \frac{1}{2} H\left(\mathcal{R}^\alpha\right) + \frac{1}{2} H\left(\mathcal{R}^{1-\alpha}\right) \right)
    + \sum_{\mathsf{m} \in \mathsf{M}} p_\mathsf{m} \left(1 - H_\alpha + \frac{1}{2} H\left(\mathcal{R}^\alpha_\mathsf{m}\right) + \frac{1}{2} H\left(R^{1-\alpha}_\mathsf{m} \right) \right),
\end{equation}
where $\mathcal{R}^\alpha = \left\{ \left(1-\alpha\right)q^\leftnode_\mathsf{m} \,|\, \mathsf{m} \in \mathsf{M} \right\} \cup \left\{ \alpha q^\rightnode_\mathsf{m} \,|\, \mathsf{m} \in \mathsf{M} \right\}$ is the set of mixed module entry rates and $\mathcal{R}^\alpha_\mathsf{m}$ is the set of mixed node visit rates in module $\mathsf{m}$, as defined in \Eqnref{eqn:derivation-r-alpha}.

\noindent
Based on \Eqnref{eqn:derivation-two-level-rewrite-alpha}, we define a first version of the bipartite map equation with varying node-type memory,
\begin{equation}
  L^\alpha\left(\mathsf{M}\right)
    = \frac{q}{2} \left( H\left(\mathcal{R}^\alpha)\right) + H\left(\mathcal{R}^{1-\alpha}\right) \right)
    + \sum_{\mathsf{m} \in \mathsf{M}} \frac{p_\mathsf{m}}{2} \left(H\left(\mathcal{R}^\alpha_\mathsf{m}\right) + H\left(\mathcal{R}^{1-\alpha}_\mathsf{m}\right)\right)
\end{equation}

\noindent
Finally, we assume that node types are fuzzy and are flipped at rate $\alpha$.
A node that is in fact a left node \emph{appears} to be a right node an $\alpha$-fraction of the time.
Similarly, a right node \emph{appears} to be a left node an $\alpha$-fraction of the time.
This means that, on average, node types are mixed and have both left and right components.
We model this with pairs: left nodes $u$ with visit rate $p_u \in \mathcal{P}^\leftnode$ have a mixed visit rate $p_u^\alpha = \left(\left(1-\alpha\right) p_u, \alpha p_u\right)$, and right nodes $v$ with visit rate $p_v \in \mathcal{P}^\rightnode$ have a mixed visit rate $p_v^\alpha = \left(\alpha p_v, \left(1-\alpha\right) p_v\right)$.

\noindent
Using mixed node visit rates, we refine our earlier definition from \Eqnref{eqn:derivation-r-alpha} and combine $\mathcal{R}^\alpha$ and $\mathcal{R}^{1-\alpha}$,
\begin{equation}
  \label{eqn:derivation-p-alpha}
  \mathcal{P}^\alpha
    = \left\{\left(\left(1-\alpha\right) p_u, \alpha p_u\right) |\, p_u \in \mathcal{P}^\leftnode \right\}
    \cup \left\{\left(\alpha p_v, \left(1-\alpha\right) p_v\right) |\, p_v \in \mathcal{P}^\rightnode\right\}.
\end{equation}
Further, all codebook usage rates become pairs, $q^\alpha = \sum_{\mathsf{m} \in \mathsf{M}} q^\alpha_\mathsf{m}$ and $p^\alpha_\mathsf{m} = q^\alpha_\mathsf{m} + \sum_{u \in \mathsf{m}^\leftnode} p^\alpha_u + \sum_{v \in \mathsf{m}^\rightnode} p^\alpha_v$ where $q^\alpha_\mathsf{m}$ is the mixed module entry and exit rate of $\mathsf{m}$ and addition works component-wise.
Since the network is bipartite and random walks alternate between left and right nodes, we have $q^\alpha = \left(\frac{q}{2}, \frac{q}{2}\right)$ and $p^\alpha_\mathsf{m} = \left(\frac{p_\mathsf{m}}{2},\frac{p_\mathsf{m}}{2}\right)$.

\noindent
Combining Eqs. \ref{eqn:derivation-r-alpha}-\ref{eqn:derivation-p-alpha}, we define the bipartite map equation with varying node-type memory,
\begin{equation}
  L^\alpha\left(M\right) = q^\alpha H\left(\mathcal{Q}^\alpha\right) + \sum_{\mathsf{m} \in \mathsf{M}} p^\alpha_\mathsf{m} H\left(\mathcal{P}^\alpha_\mathsf{m}\right),
\end{equation}
where $\mathcal{Q}^\alpha = \left\{q^\alpha_\mathsf{m} \,|\, \mathsf{m} \in \mathsf{M} \right\}$ is the set of mixed module entry rates and $\mathcal{P}^\alpha_\mathsf{m} = \left\{q^\alpha_\mathsf{m}\right\} \cup \left\{p^\alpha_u \,|\, u \in \mathsf{m}^\leftnode\right\} \cup \left\{p^\alpha_v \,|\, v \in \mathsf{m}^\rightnode\right\}$ is the set of mixed node visit rates in module $\mathsf{m}$, including module exits.

\clearpage
\section{Community landscapes}
\label{appendix:plots}
This appendix contains short descriptions of all used networks and plots with the results from our analyses.
The solid and dashed blue lines show the extra compression of the best hierarchical and two-level partitions, respectively, and the solid and dashed orange lines show the effective module size of the best hierarchical and two-level partitions, respectively.
Where it was useful to reveal more details, we plotted the effective module size on a logarithmic scale.

\begin{figure}[ht!]
  \subfloat[]{%
    \includegraphics[width=.25\columnwidth]{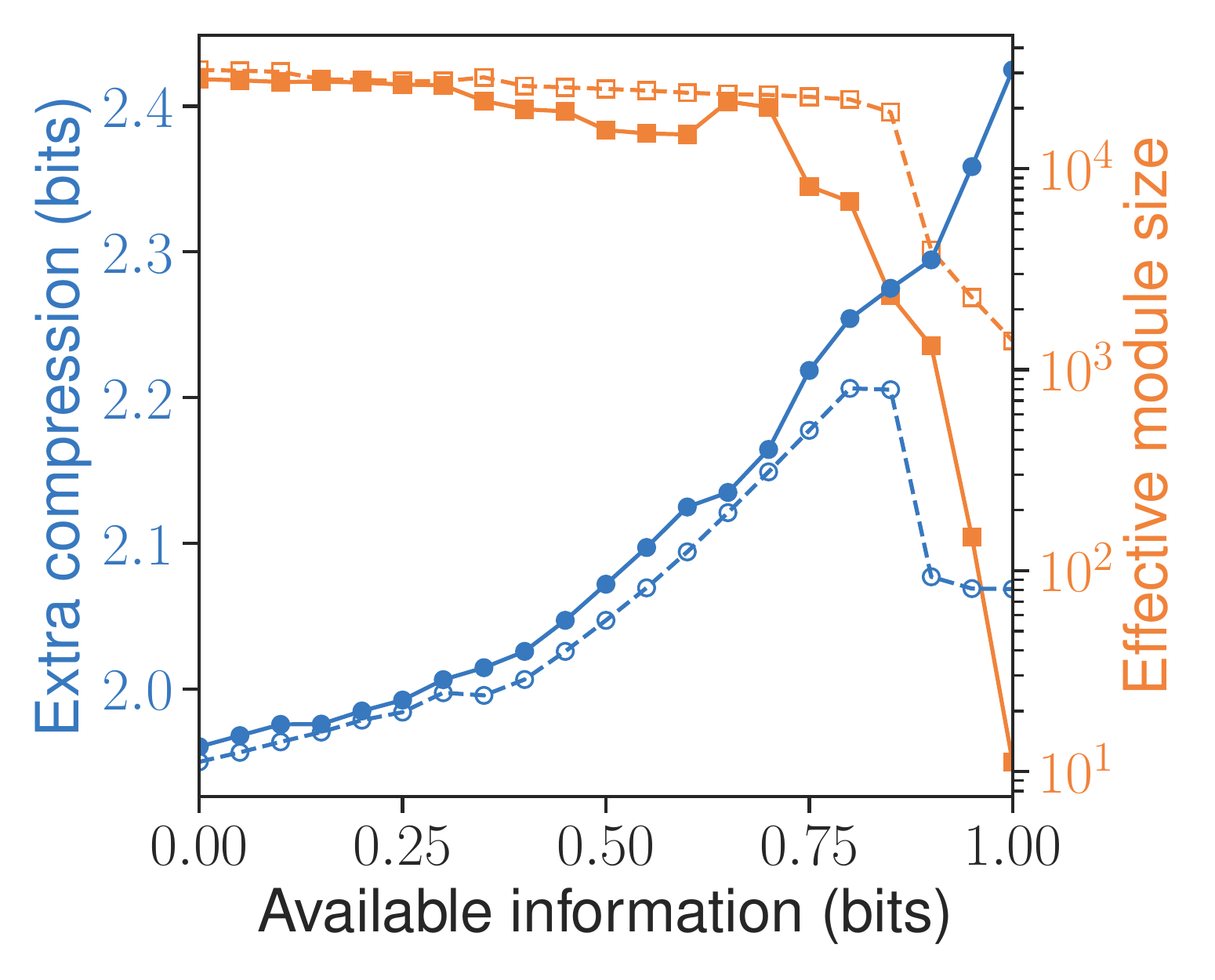}%
  }\hfill
  \subfloat[]{%
    \includegraphics[width=.25\columnwidth]{lastfm-song-sweep}%
  }\hfill
  \subfloat[]{%
    \includegraphics[width=.25\columnwidth]{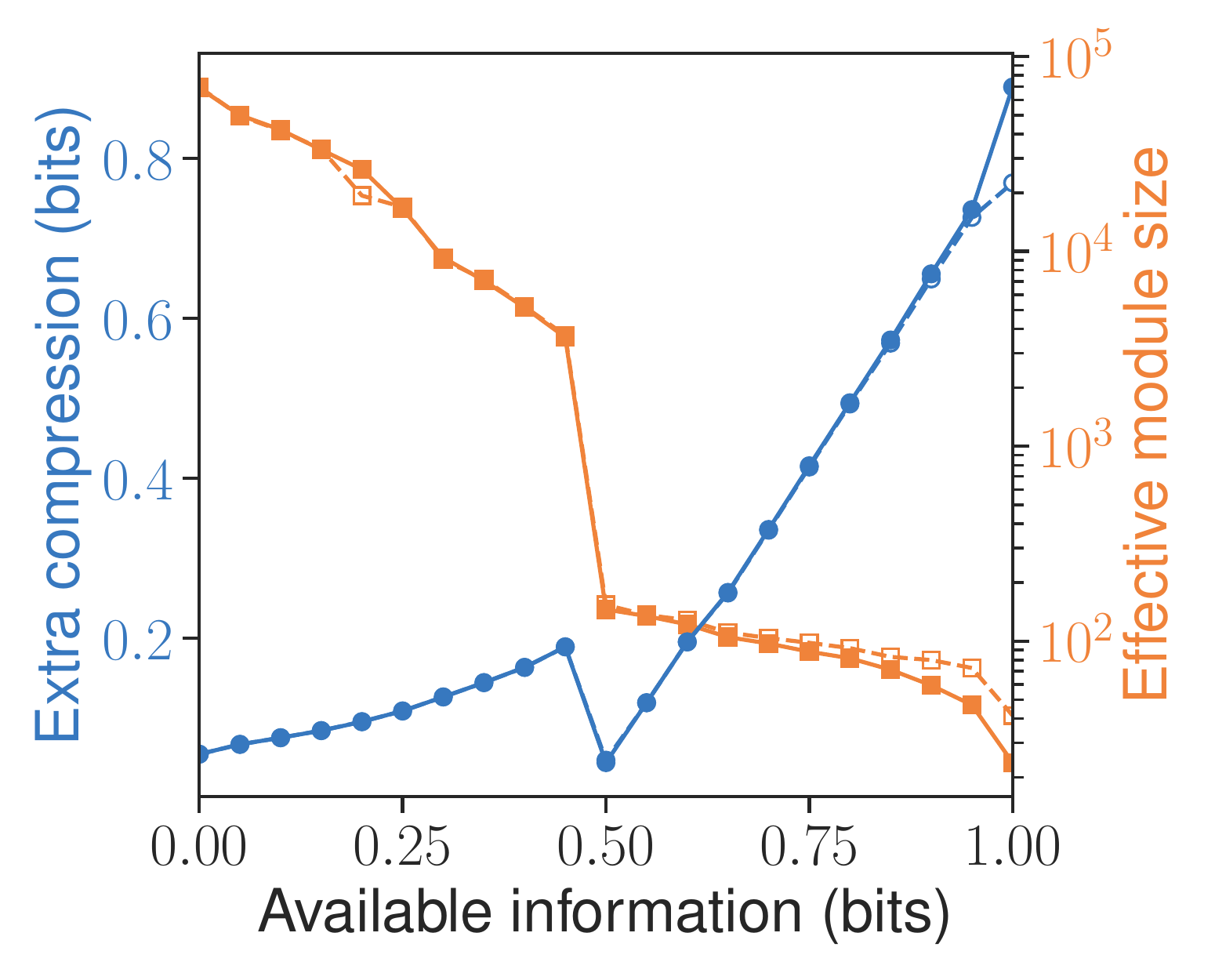}%
  }\hfill
  \subfloat[]{%
    \includegraphics[width=.25\columnwidth]{IMDb-actor-movie-sweep}%
  }

  \subfloat[]{%
    \includegraphics[width=.25\columnwidth]{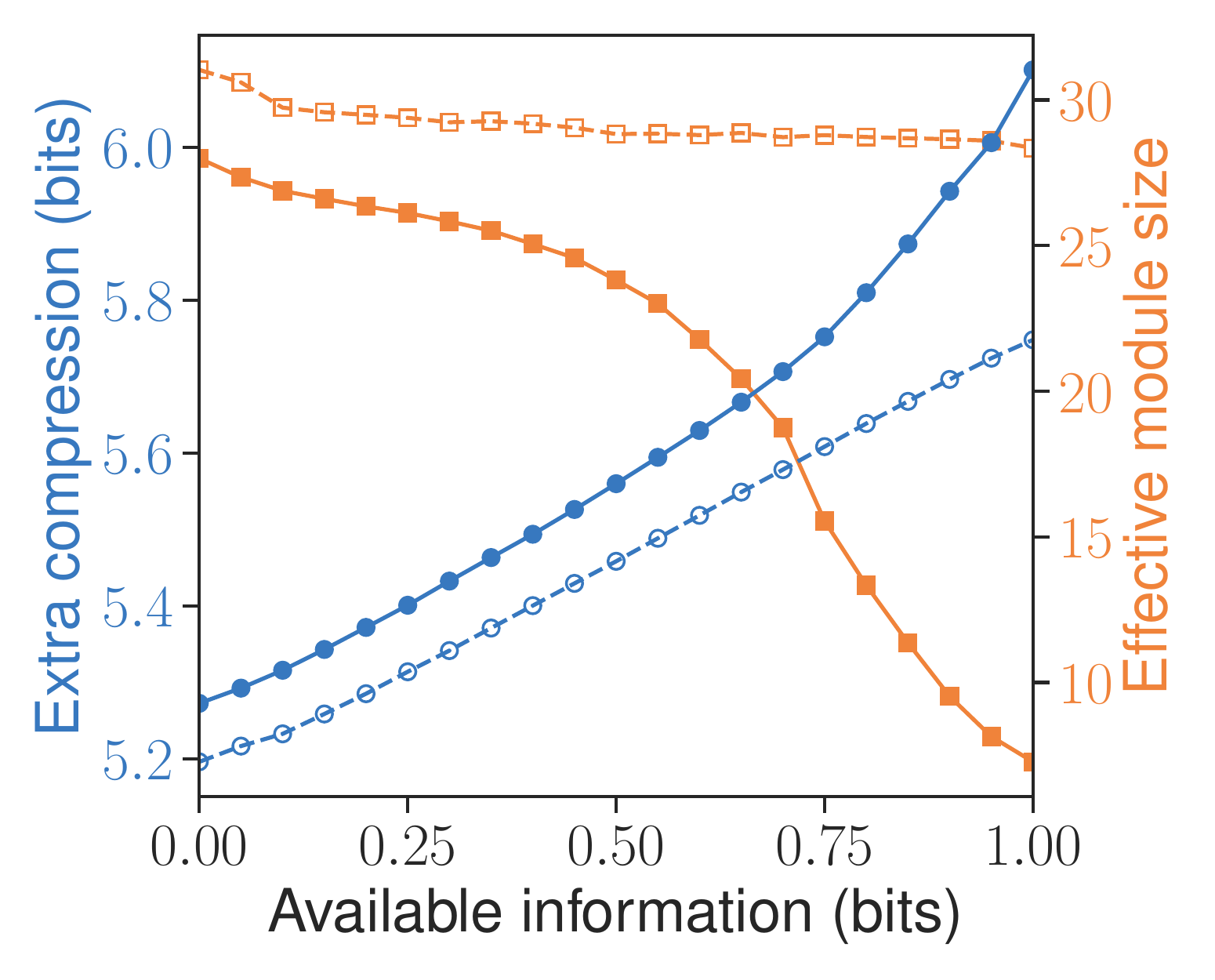}%
  }\hfill
  \subfloat[]{%
    \includegraphics[width=.25\columnwidth]{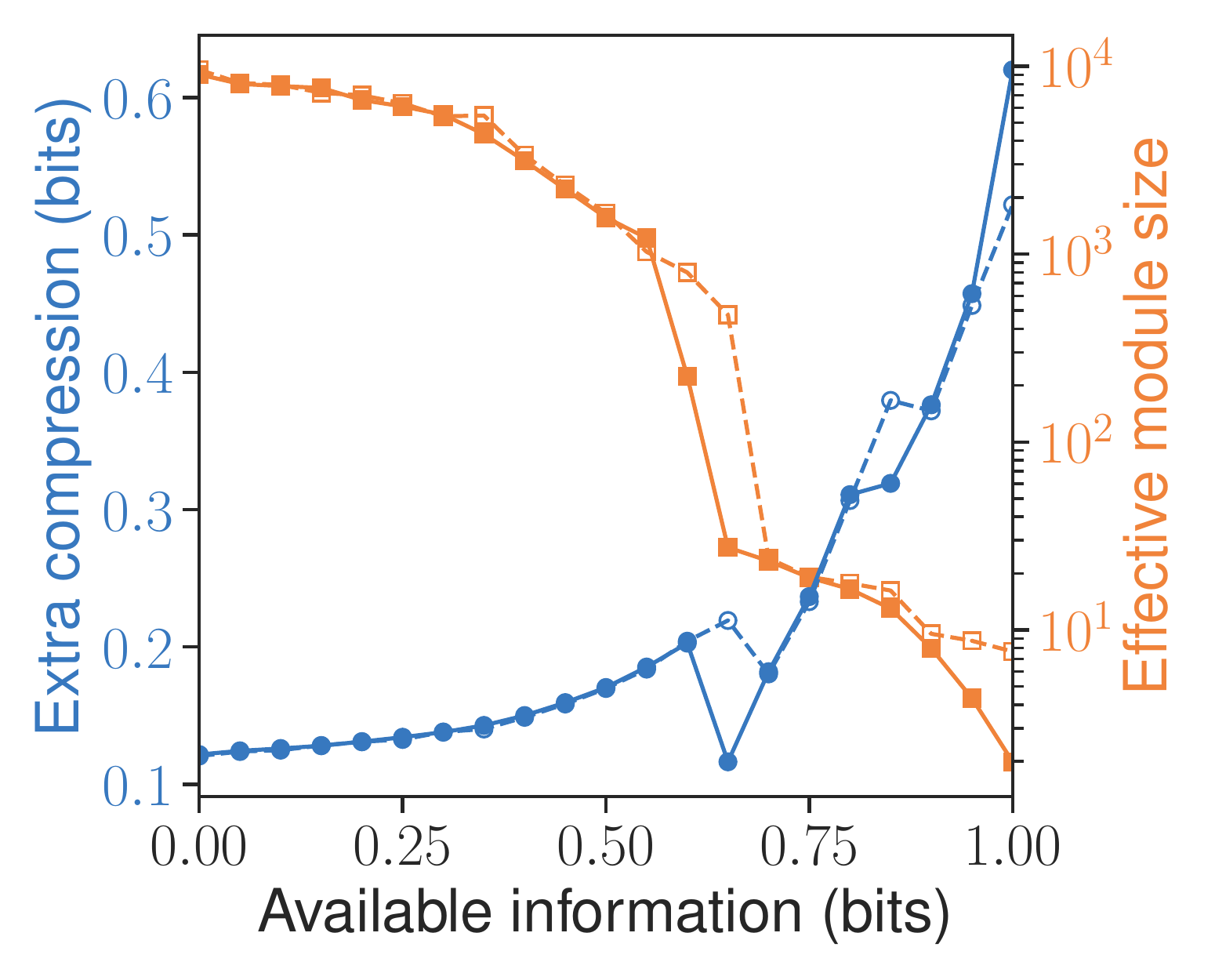}%
  }\hfill
  \subfloat[]{%
    \includegraphics[width=.25\columnwidth]{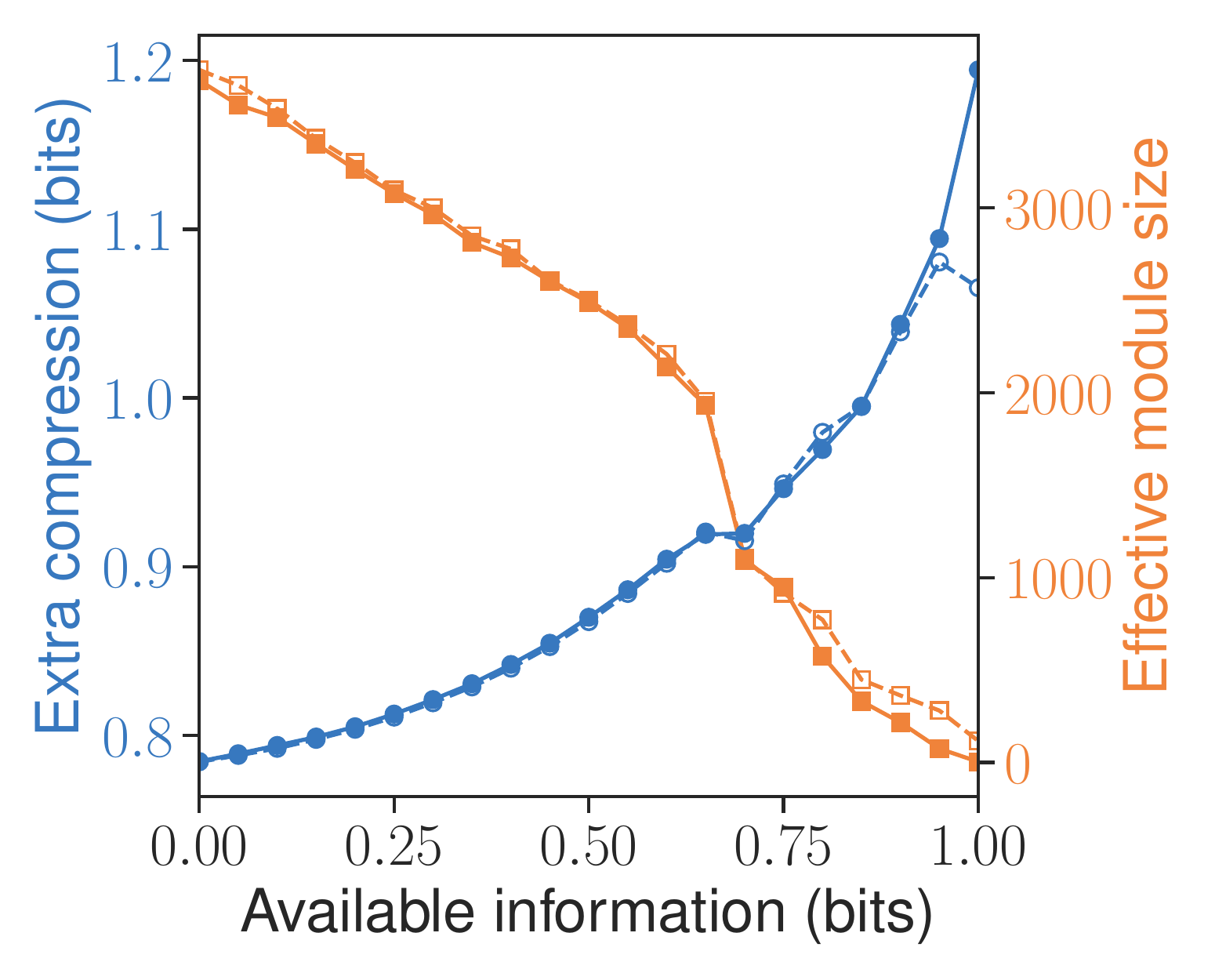}%
  }\hfill
  \subfloat[]{%
    \includegraphics[width=.25\columnwidth]{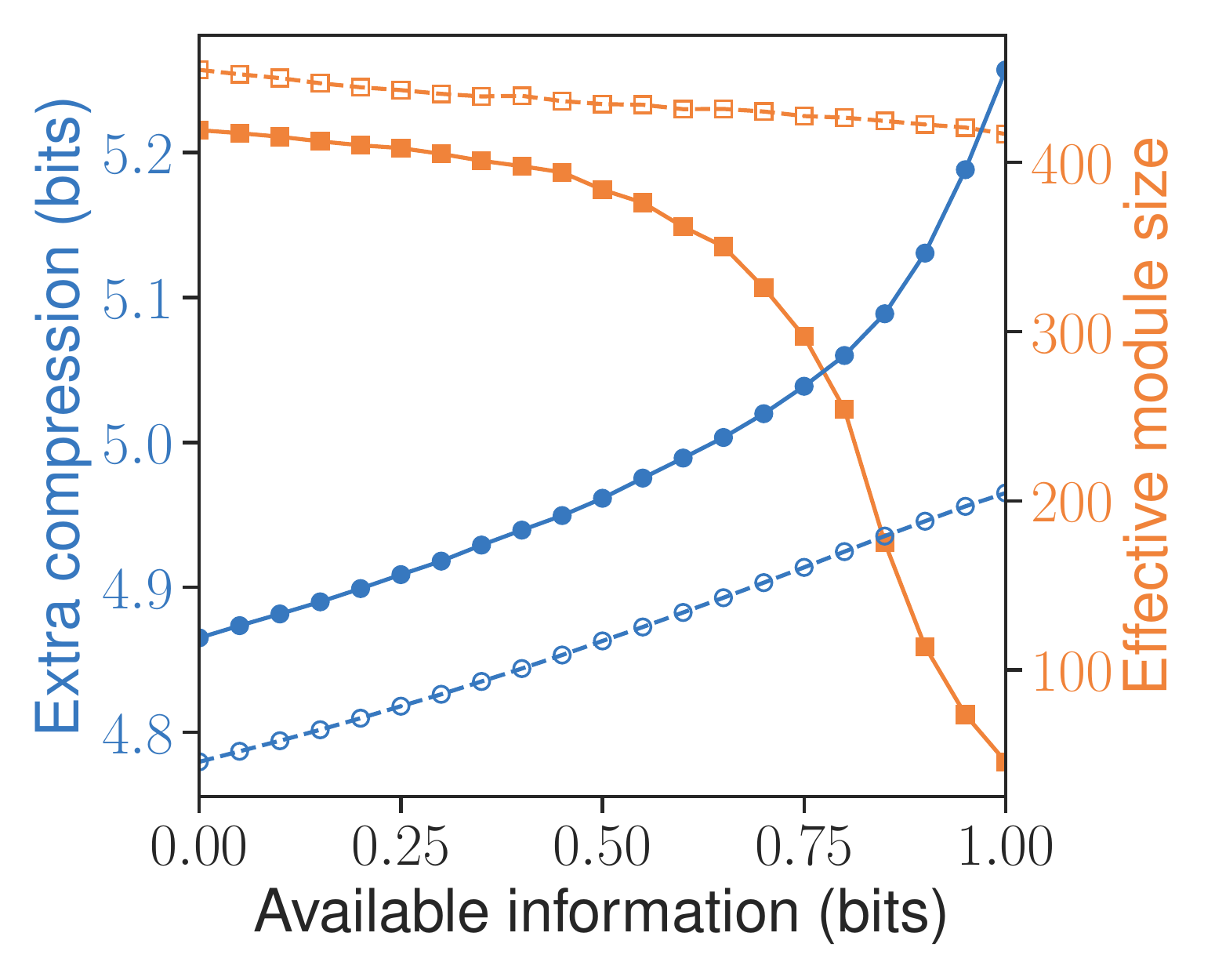}%
  }

  \subfloat[]{%
    \includegraphics[width=.25\columnwidth]{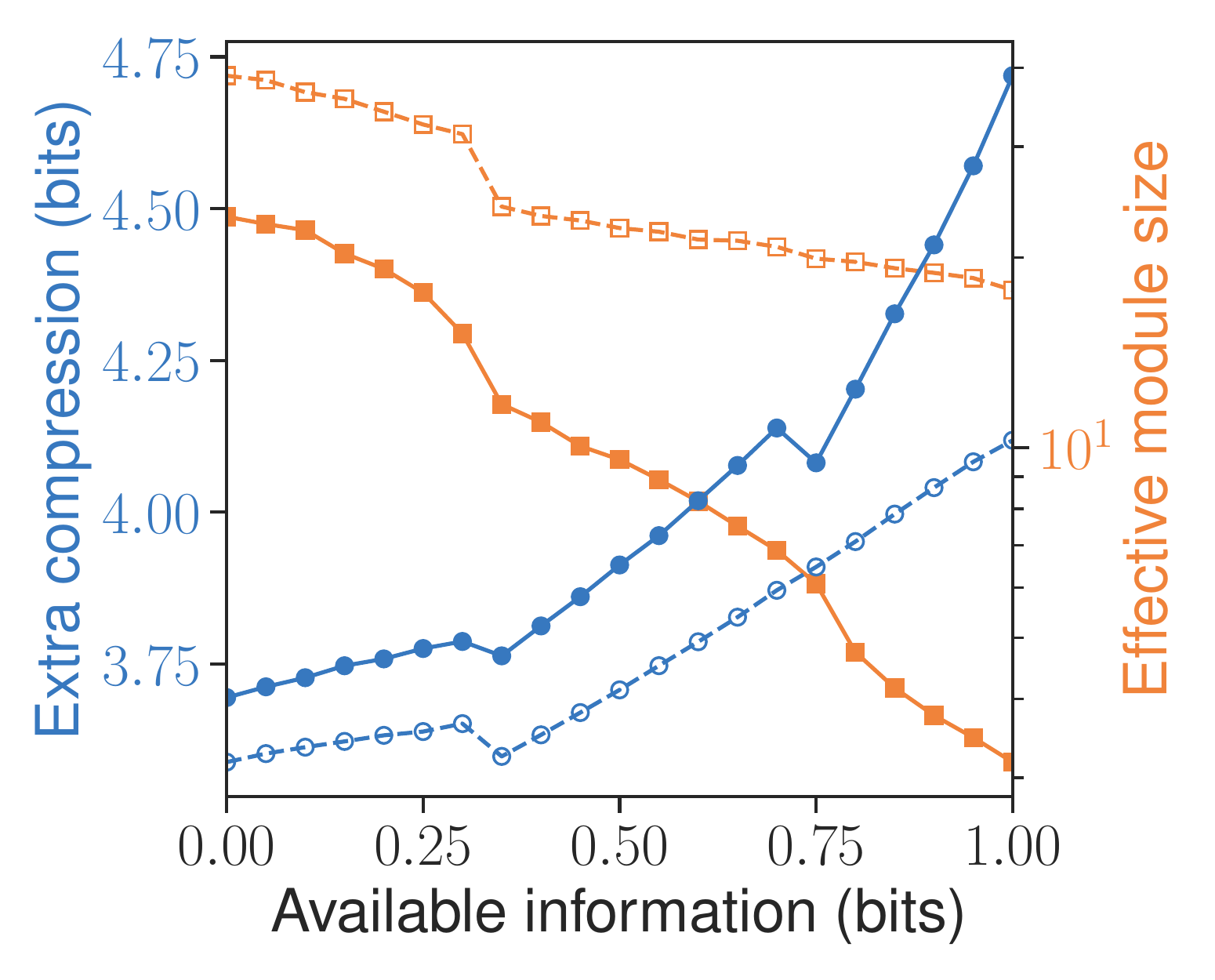}%
  }\hfill
  \subfloat[]{%
    \includegraphics[width=.25\columnwidth]{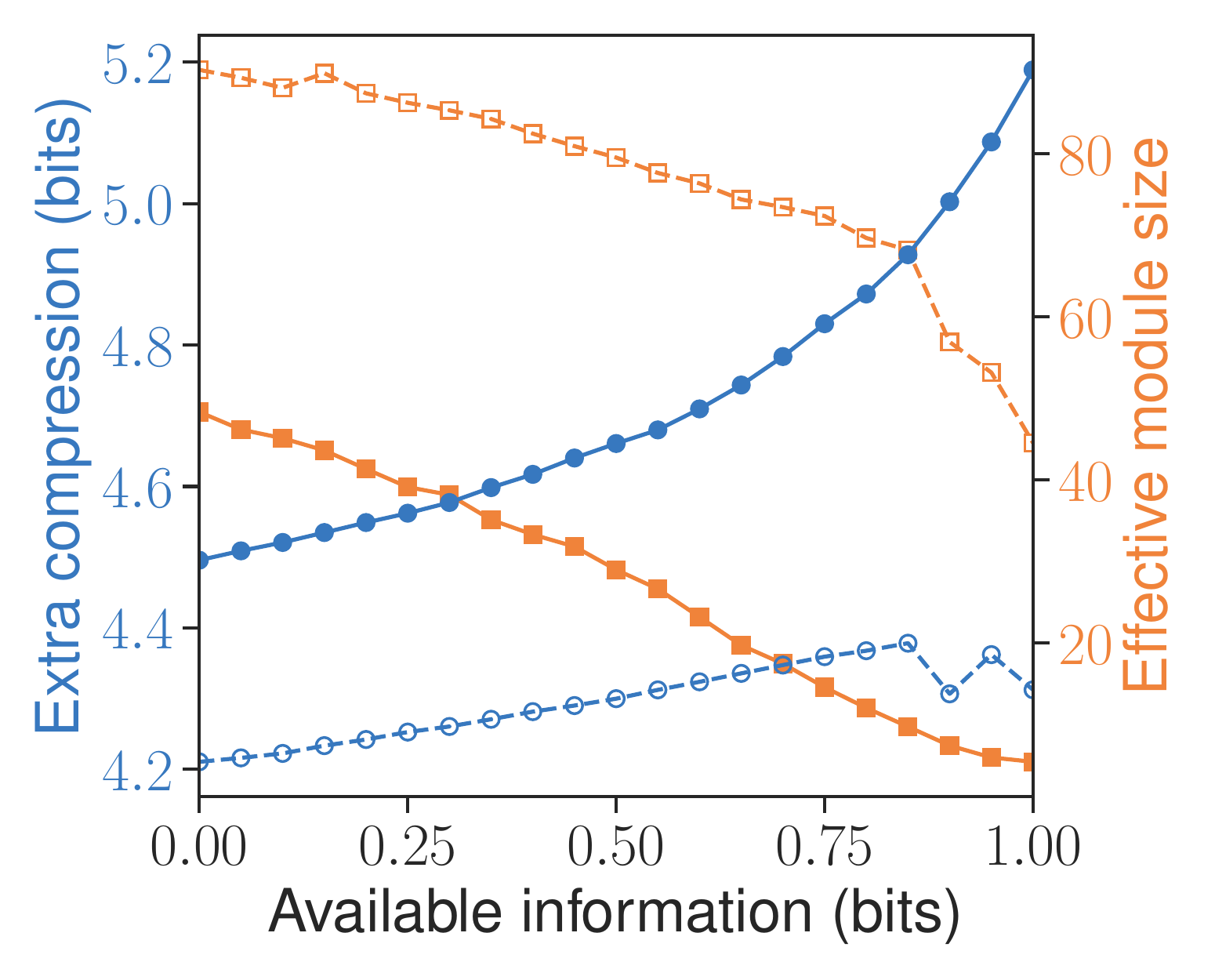}%
  }\hfill
  \subfloat[]{%
    \includegraphics[width=.25\columnwidth]{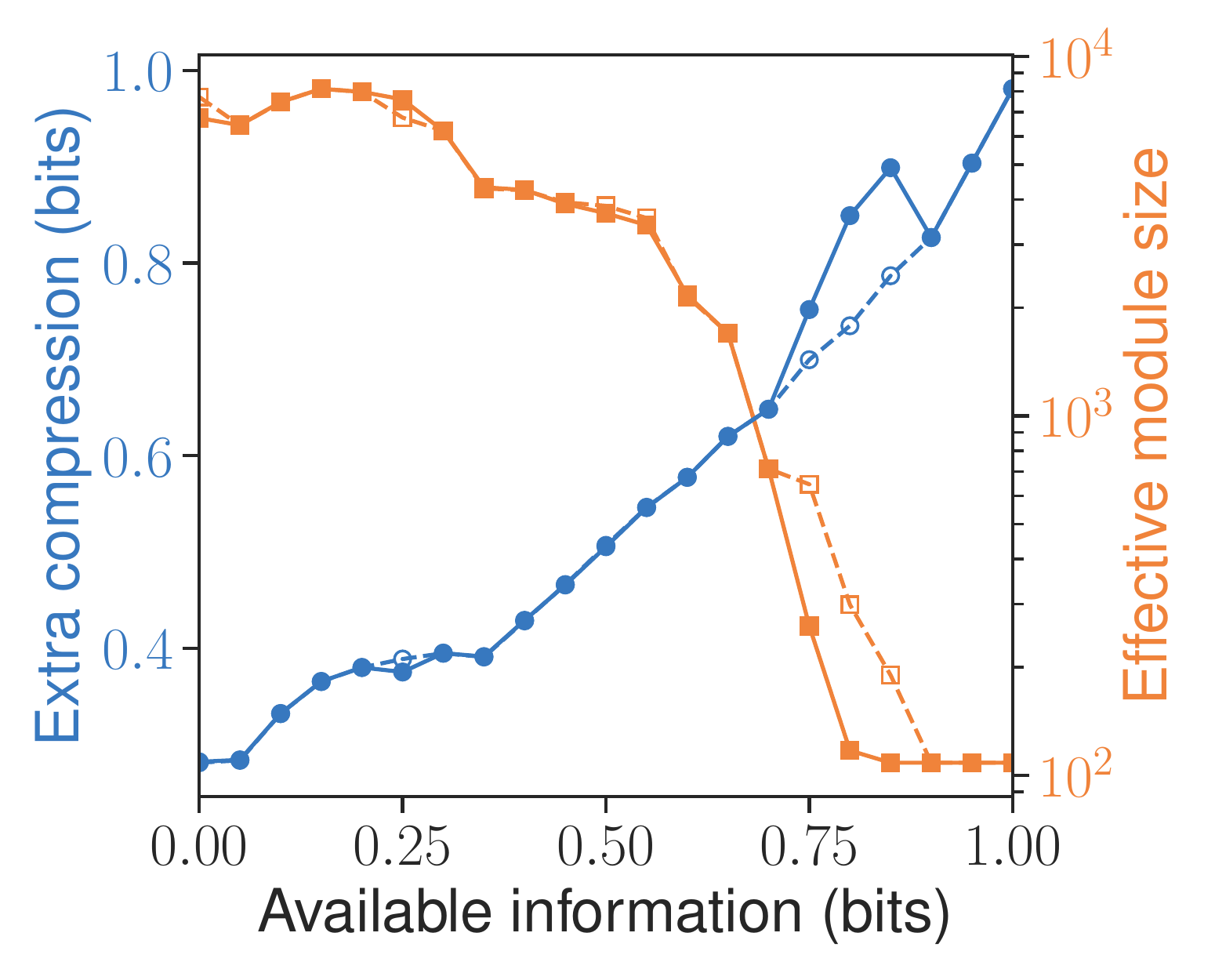}%
  }\hfill
  \subfloat[]{%
    \includegraphics[width=.25\columnwidth]{LVHK-sweep}%
  }

  \subfloat[]{%
    \includegraphics[width=.25\columnwidth]{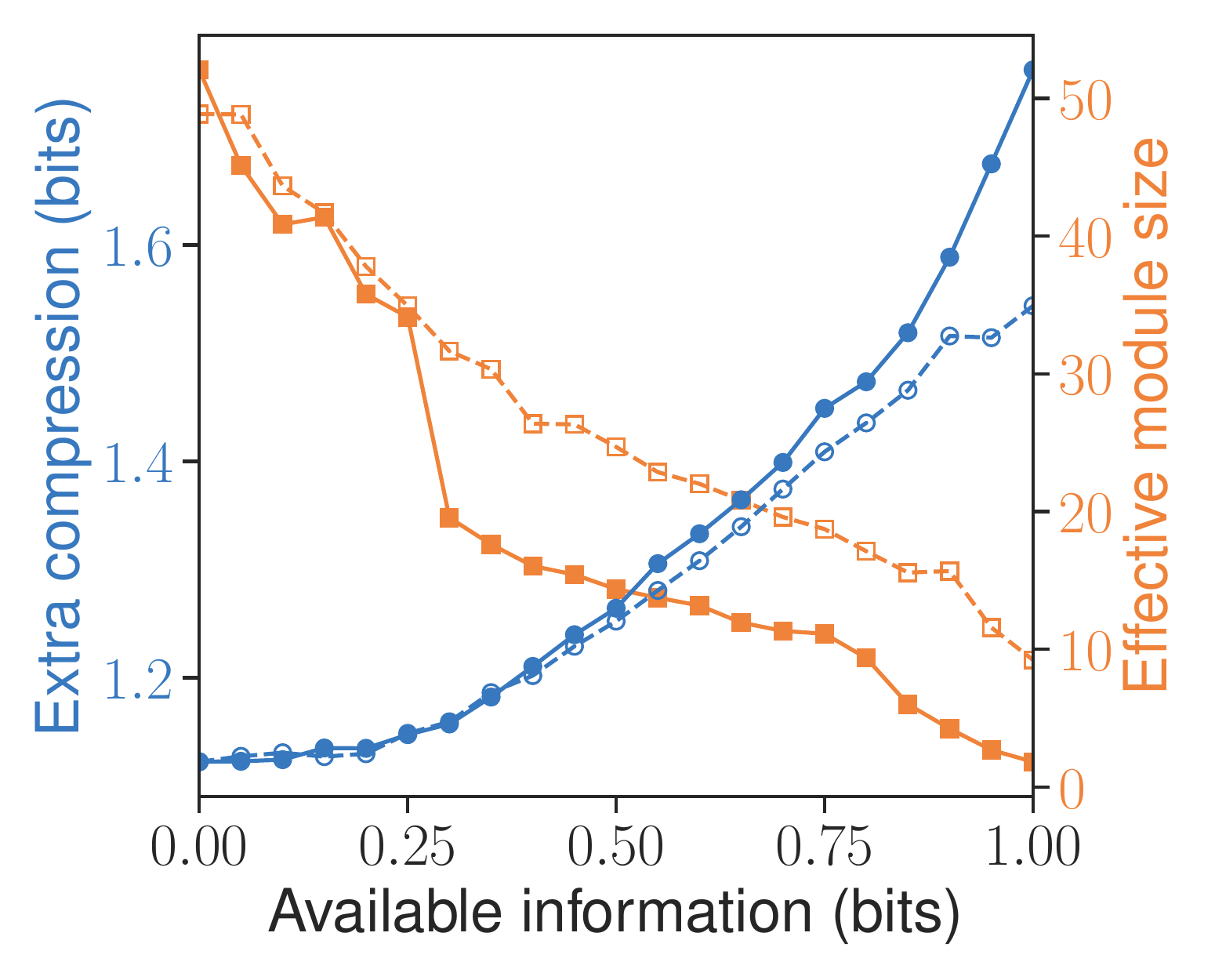}%
  }\hfill
  \subfloat[]{%
    \includegraphics[width=.25\columnwidth]{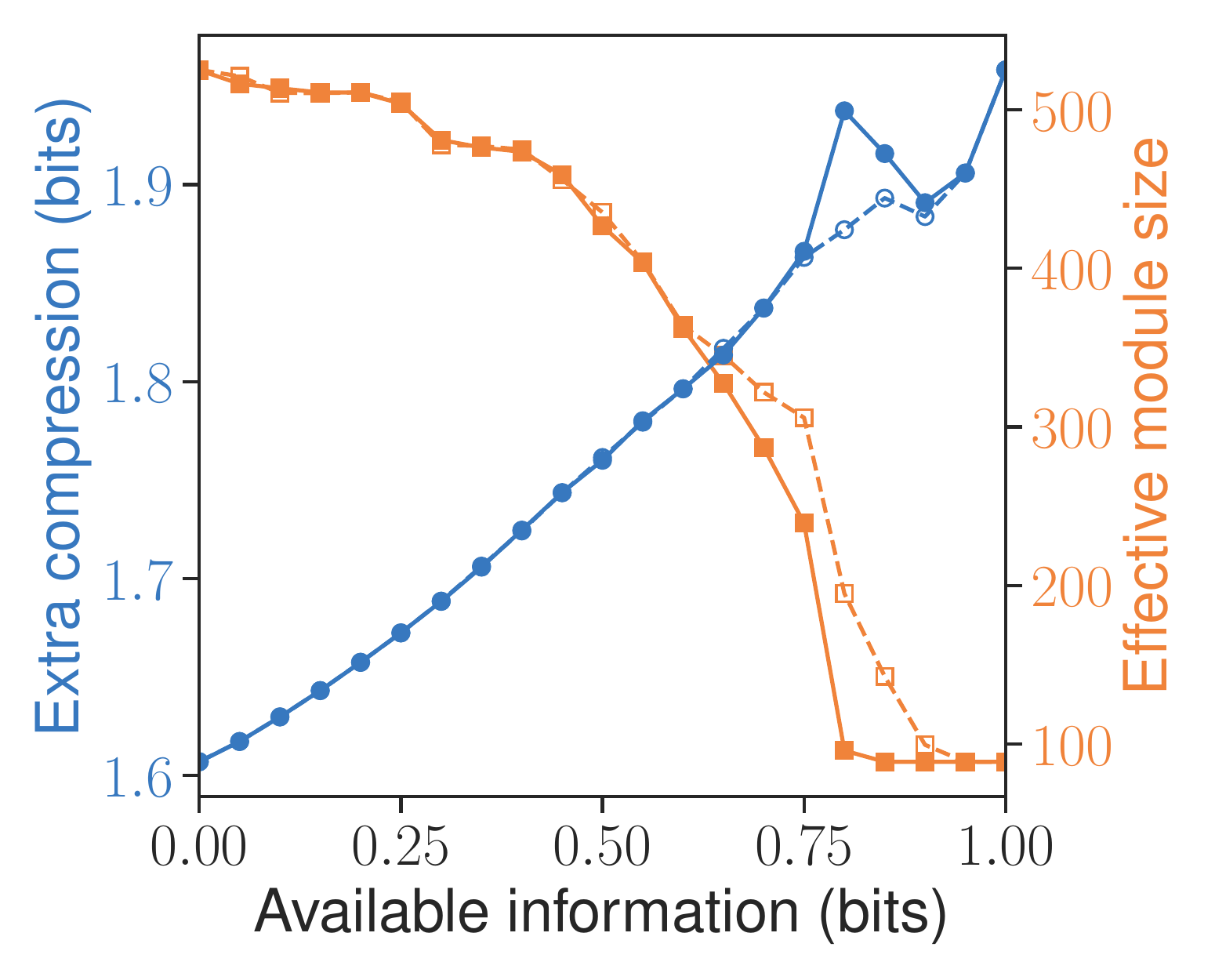}%
  }\hfill
  \subfloat[]{%
    \includegraphics[width=.25\columnwidth]{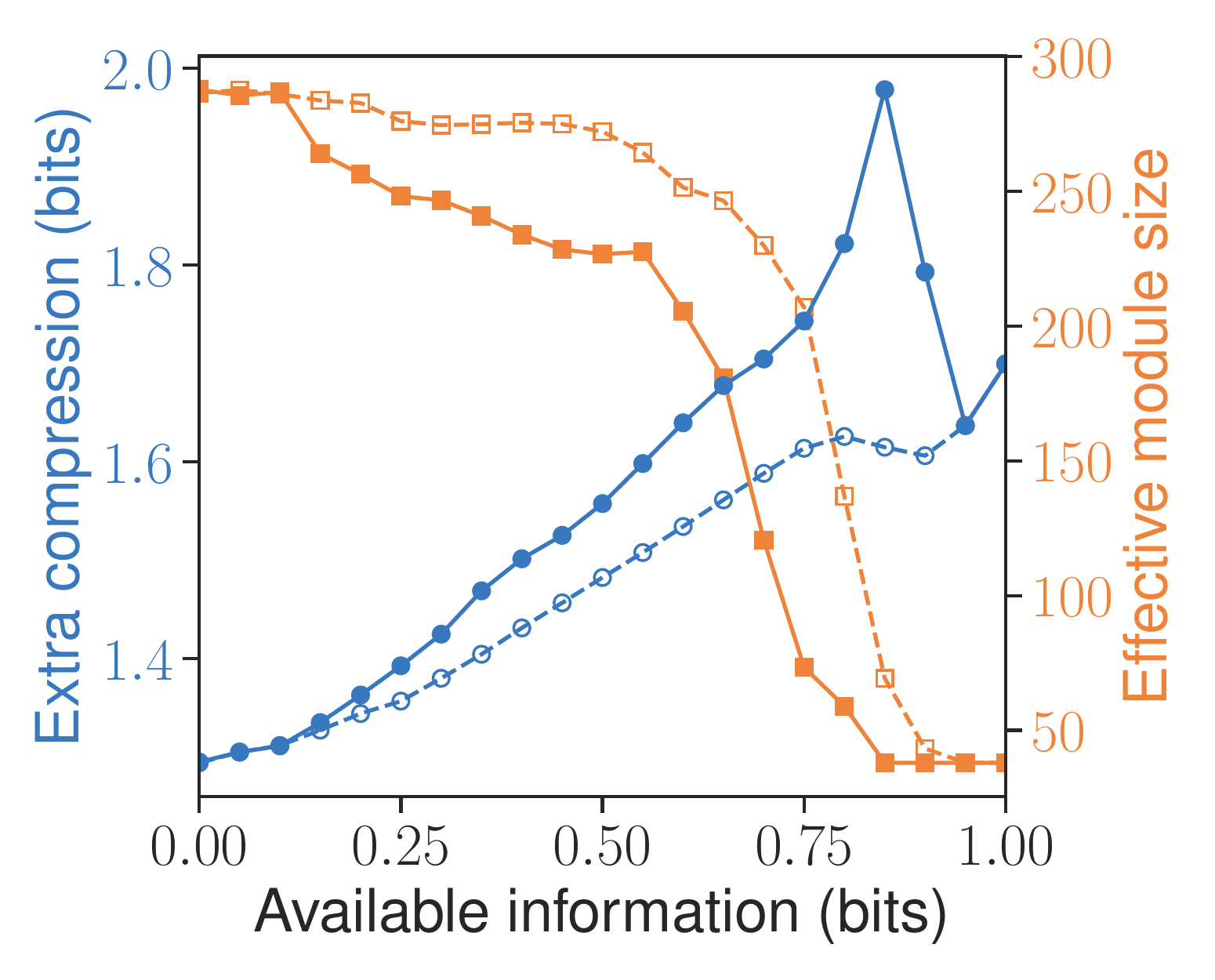}%
  }\hfill
  \subfloat[]{%
    \includegraphics[width=.25\columnwidth]{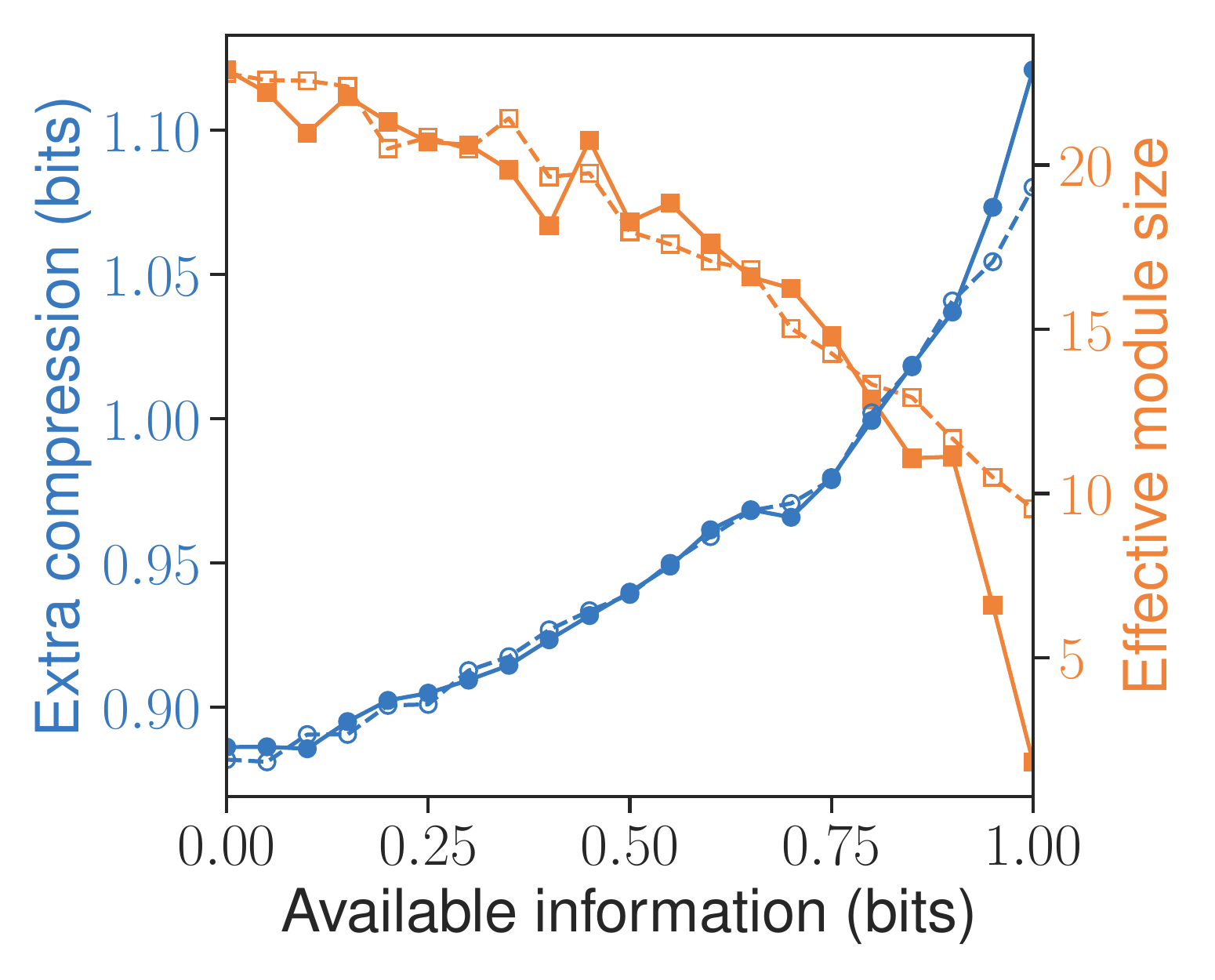}%
  }\hfill
\end{figure}

\begin{figure}[ht!]\ContinuedFloat
  \subfloat[]{%
    \includegraphics[width=.25\columnwidth]{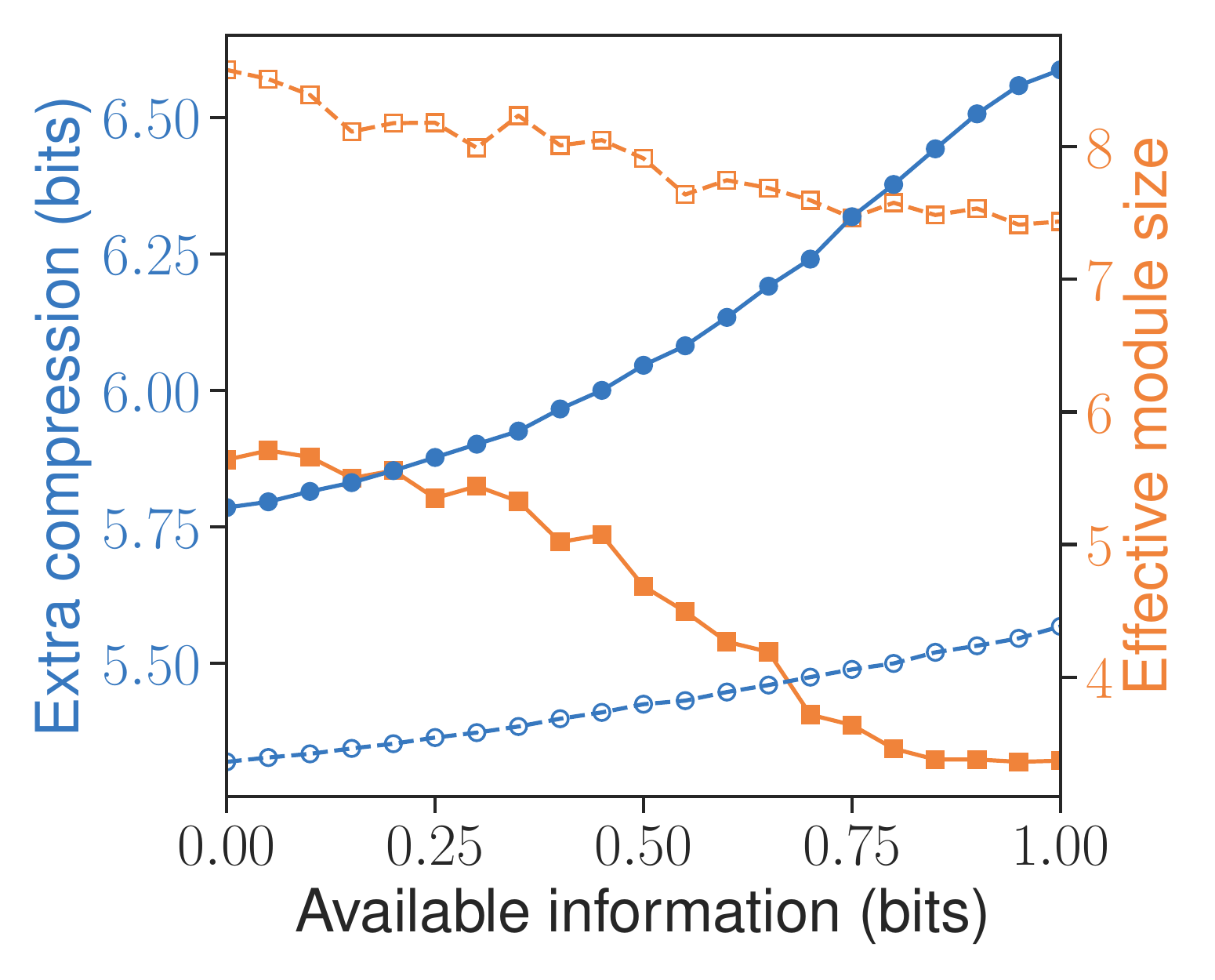}%
  }\hfill
  \subfloat[]{%
    \includegraphics[width=.25\columnwidth]{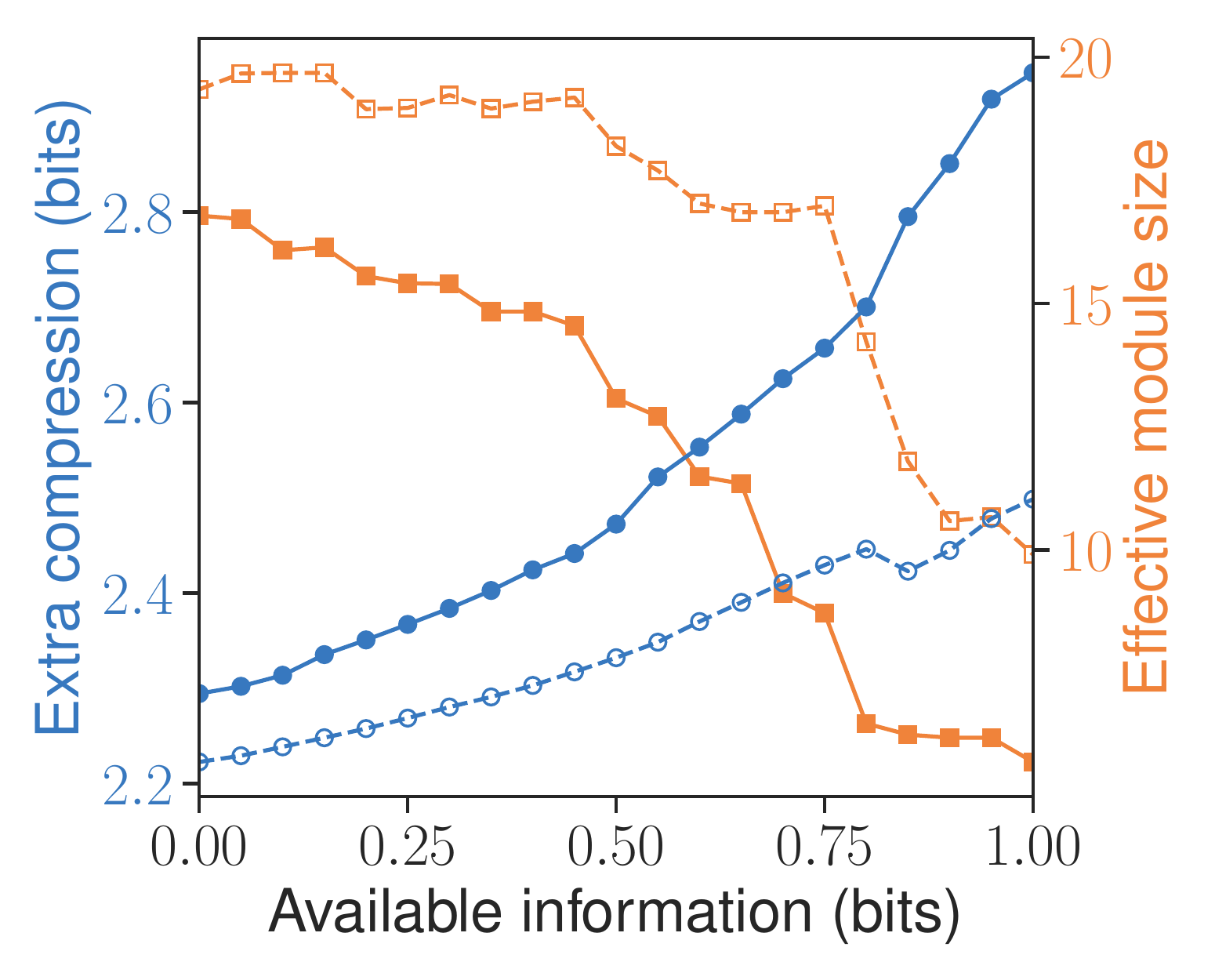}%
  }\hfill
  \subfloat[]{%
    \includegraphics[width=.25\columnwidth]{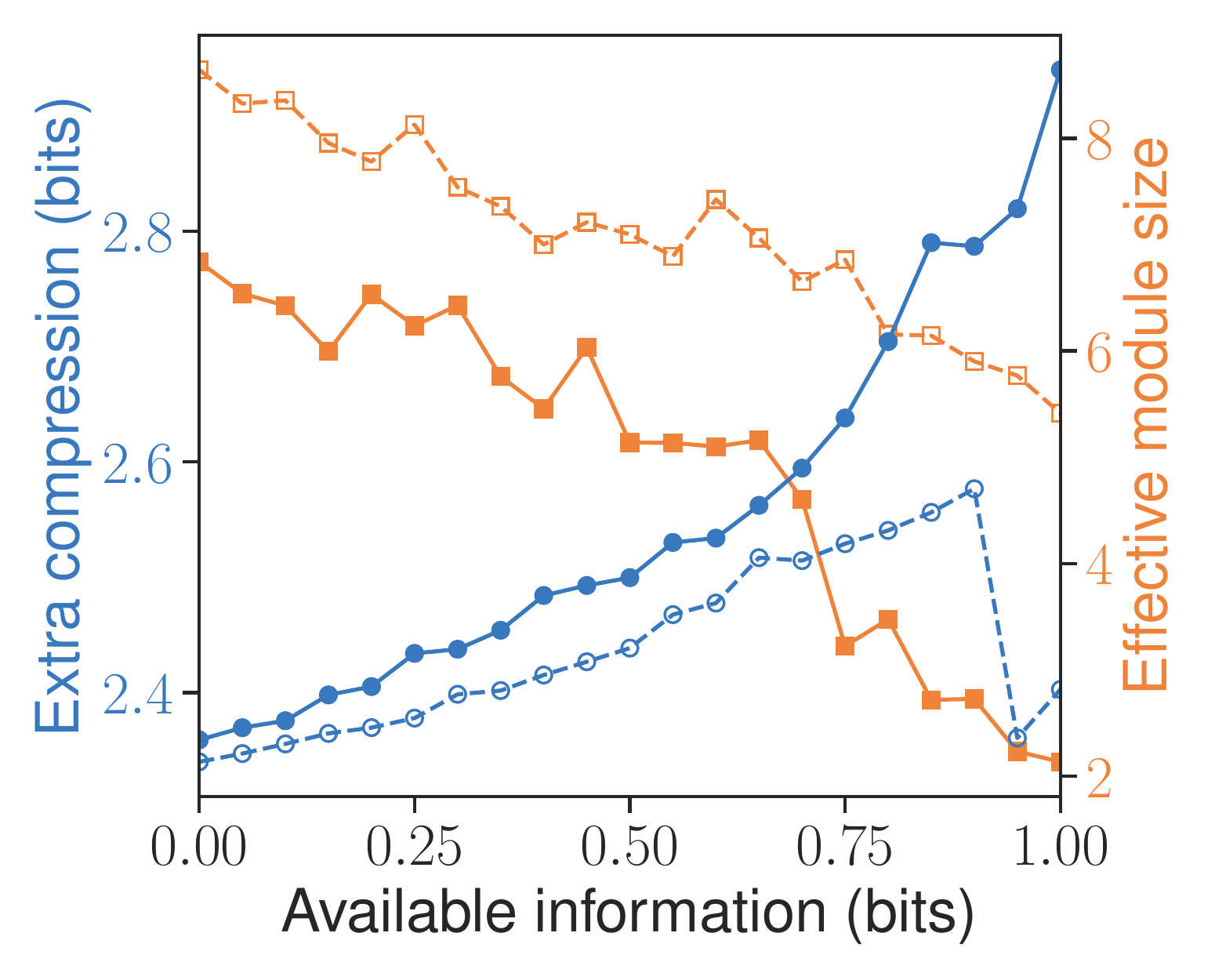}%
  }\hfill
  \subfloat[]{%
    \includegraphics[width=.25\columnwidth]{Vazquez-Simberloff-arroyo-goye-weighted-sweep}%
  }

  \subfloat[]{%
    \includegraphics[width=.25\columnwidth]{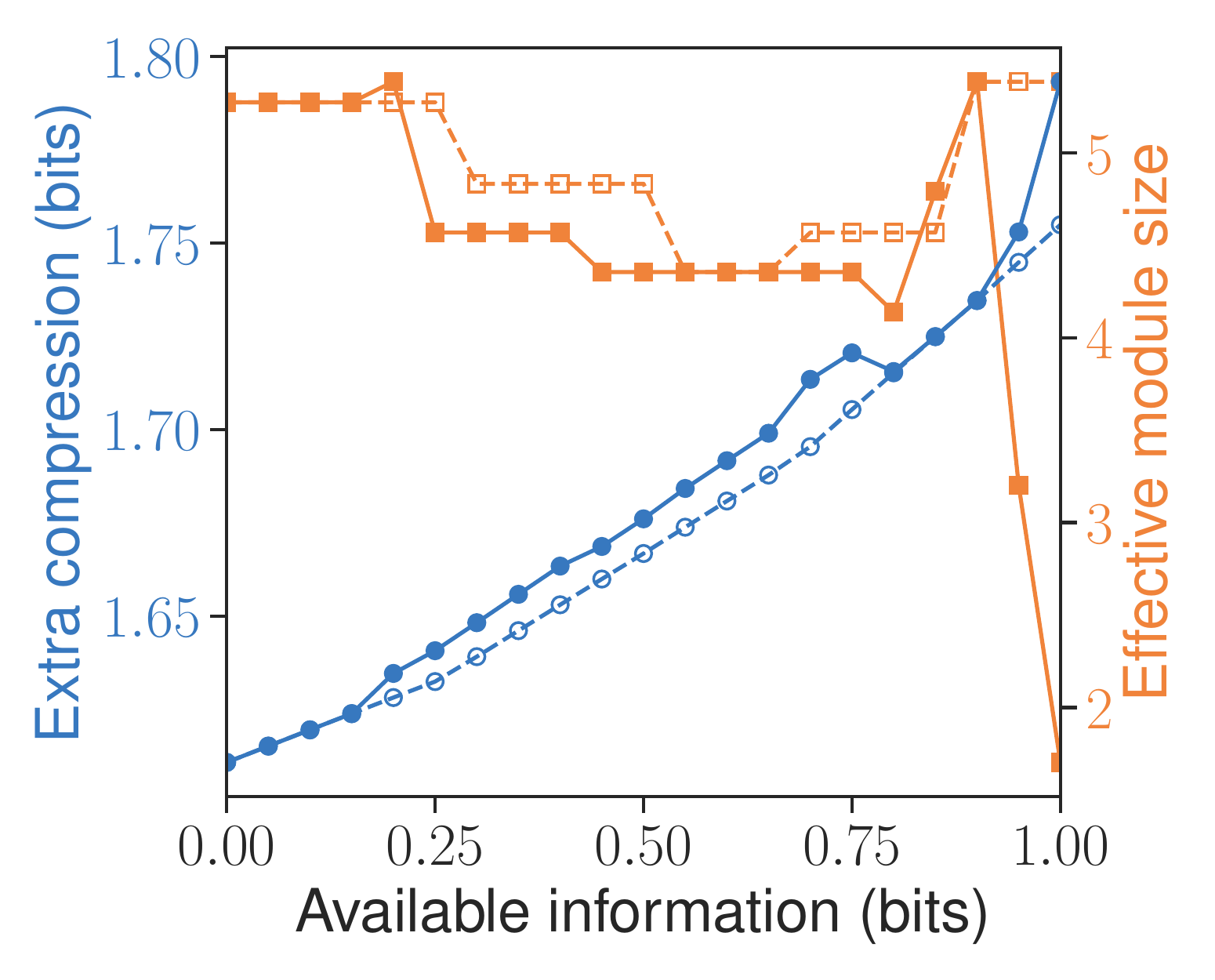}%
  }
  \caption{
    (a) Wiktionary (en): left nodes represent authors, right nodes represent articles on English Wiktionary. An edge connects authors to the articles they have authored.
    (b) Last.fm user-song: left nodes represent users, right nodes represent songs. Edges connect users to the songs they have listened to.
    (c) Wikipedia excellent: left node represent excellent articles on Wikipedia, right nodes represent words. An edge connects an article to a word if it contains it.
    (d) IMDb actor-movie: left nodes represent actors, right nodes represent movies. An edges connects actors to those movies they have played in.
    (e) Stack Overflow user-post: left nodes represent users, right nodes represent posts. An edge connects users to those posts they have marked as a favorite.
    (f) Reuters story-word: left nodes represent stories in the Reuters Corpus, Volume 1, right nodes represent words. An edge connects a story to a word if it contains it.
    (g) Wiktionary (de): left nodes represent authors, right nodes represent articles on German Wiktionary. An edge connects authors to the articles they have authored.
    (h) Linux kernel mailing list: left nodes represent users, right nodes represent threads in the linux kernel mailing list. An edge connects user to those threads where they contribute.
    (i) GitHub user-project: left nodes represent users, right nodes represent projects. An edge connects users to those projects where they are a member.
    (j) YouTube user-group: left nodes represent users, right nodes represent groups. An edge connects users to the groups where they are a member.
    (k) APSMM conference: left nodes represent scientists, right nodes represent editions of the APSMM conference. Edges connect scientists to the editions of the conference they have attended.
    (l) LVHK Meetup: left nodes represent persons, right nodes represent events of the VegasHikers group on Meetup. Edges connect persons to those events they have attended.
    (m) PGHF Meetup: left nodes represent persons, right nodes represent events of the pittsburgh-free group on Meetup. Edges connect persons to those events they have attended.
    (n) SIAM conference: left nodes represent scientists, right nodes represent editions of the SIAM conference. Edges connect scientists to the editions of the conference they have attended.
    (o) NIPS conference: left nodes represent scientists, right nodes represent editions of the NIPS conference. Edges connect scientists to the editions of the conference they have attended.
    (p) UC Irvine forum: left nodes represent users, right nodes represent topics in the UC Irvine online forum. An edge connects users to those topics where they have made a post.
    (q) Norwegian directors: left nodes represent directors, right nodes represent Norwegian companies. Edges connect persons to the companies where they are member of the board of directors.
    (r) Virus-host interactome: left nodes represent virus proteins, right nodes represent host proteins. An edge connects virus proteins to those host proteins they interact with.
    (s) Scottish directors: left nodes represent directors, right nodes represent Scottish companies. Edges connect directors to the companies where they are member of the board of directors.
    (t) Arroyo Goye pollinator-plant: left nodes represent pollinators, right nodes represent plant species. An edge connects pollinators to the plants they pollinate.
    (u) Fonseca Ganade ant-plant: left nodes represent ant species, right nodes represent plant species. Edges connect ant species to those plant species that they use as a source of food or housing.
    }
\end{figure}


\end{document}